\documentclass[3p, final, times]{elsarticle}

\usepackage{amssymb}

\usepackage[T1]{fontenc}
\usepackage{tabularx}
\usepackage{adjustbox}
\usepackage{algpseudocode}
\usepackage{numprint} 
\usepackage{natbib}

\usepackage{xspace}
\newcommand*{\eg}{e.g.\@\xspace}
\newcommand*{\ie}{i.e.\@\xspace}
\newcommand{\norm}[1]{\left\lVert#1\right\rVert}
\makeatletter
\newcommand*{\etc}{%
	\@ifnextchar{.}%
	{etc}%
	{etc.\@\xspace}%
}
\usepackage{caption} 
\usepackage[caption=false]{subfig}
\usepackage{graphicx,amssymb}
\usepackage{url}
\usepackage{hyperref}
\usepackage{amsmath}
\usepackage{multirow}
\usepackage{rotating}
\usepackage{booktabs, multirow} 
\usepackage{changepage,threeparttable} 
\usepackage[table]{xcolor} 
\usepackage{nccmath}
\usepackage{amsmath} 
\usepackage{diagbox}

\newcommand{\comment}[1]{}

\usepackage{algorithm}
\usepackage{algpseudocode}
\usepackage{wrapfig}

\begin{document}\sloppy
	
	\begin{frontmatter}
		
		
		
		\title{A Multi-Agent Adaptive Deep Learning Framework for  Online Intrusion Detection}
		
		
		\author[sharif]{Mahdi~Soltani}
		\ead{mahdi@ce.sharif.edu}
		\author[sharif]{Khashayar~Khajavi}
		\ead{khajaviii@ce.sharif.edu}
		\author[sharif]{Mahdi~Jafari~Siavoshani\corref{cor1}}
		\ead{mjafari@sharif.edu}
		\author[sharif]{Amir~Hossein~Jahangir\corref{cor1}}
		\ead{jahangir@sharif.edu}
		
		\address[sharif]{Department of Computer Engineering, Sharif University of Technology, Tehran, Iran}
		
		\cortext[cor1]{Corresponding author.}
		
		\begin{abstract}
			The network security analyzers use intrusion detection systems (IDSes) to distinguish malicious traffic from benign ones. 
			The deep learning-based (DL-based) IDSes are proposed to auto-extract high-level features and eliminate the time-consuming and costly signature extraction process. 
			However, this new generation of IDSes still suffers from a number of challenges.
			One of the main issues of an IDS is facing traffic concept drift which manifests itself as new (\ie, zero-day) attacks, in addition to the changing behavior of benign users/applications.
			Furthermore, a practical DL-based IDS needs to be conformed to a distributed architecture to handle big data challenges.

			In this paper, we propose a framework for adapting DL-based models to the changing attack/benign traffic behaviors, considering a more practical scenario (\ie, online adaptable IDSes).
			This framework employs continual deep anomaly detectors in addition to the federated learning approach to solve the above-mentioned challenges. Furthermore, the proposed framework implements sequential packet labeling for each flow, which provides an attack probability score for the flow by gradually observing each flow packet and updating its estimation.
			We evaluate the proposed framework by employing different deep models (including CNN-based and LSTM-based) over the CIC-IDS2017 and CSE-CIC-IDS2018 datasets. 
			Through extensive evaluations and experiments, we show that the proposed distributed framework is well adapted to the traffic concept drift. More precisely, our results indicate that the CNN-based models are well suited for continually adapting to the traffic concept drift (\ie, achieving an average detection rate of above 95\% while needing just 128 new flows for the updating phase), and the LSTM-based models are a good candidate for sequential packet labeling in practical online IDSes (\ie, detecting intrusions by just observing their first 15 packets).	
		\end{abstract}
		
		
		
		
		
		
		
		
	\end{frontmatter}
	
	
\section{Introduction}
\label{intro}
Nowadays, the growth of cyber threats highlights the importance of security devices such as intrusion detection systems (IDSes). The network security analyzers use IDSes to monitor the network data, analyze them, and detect any kind of intrusions. There are mainly two categories of intrusion detectors: signature-based and machine learning-based (ML-based) \cite{labonne2020anomaly}.

The main advantage of ML-based IDSes over signature-based ones is the absence of the costly and time-consuming signature extraction process in the former. Consequently, ML-based IDSes, especially deep learning ones, are considered as the new generation of IDS devices. The ability of deep learning-based (DL-based) IDSes to auto-extract high-level features and classify different attack/benign traffic flows is their main advantage compared to the traditional ML-based IDSes. Moreover, due to the high-dimensional processing ability of DL models, the DL-based IDSes are good candidates for inspecting traffic content, as suggested in the recently proposed Deep Intrusion Detection (DID) framework \cite{soltani2020content}.

Many studies in the literature apply deep learning methods in offline IDSes \cite{thakkar2021review}\cite{soltani2021adaptable}. Nevertheless, in this paper, we focus on simultaneously adapting DL-based IDSes for the following three practical challenges of online intrusion detection.

The first challenge is related to the continuous adaptability of a DL-based IDS to an organization's traffic since both attack and benign traffic patterns might encounter concept drift with the passage of time. 
For example, switching between semester and vacation times in the universities, adding new services to the web servers, and the emergence of new popular applications and protocols are examples of changing the content and behavior of benign user/traffic over time.
Moreover, the characteristics and content of attack traffics are also changing continuously. 
This is due to the fact that the number of revealed vulnerabilities is increasing \cite{vunlerability2020}, and additionally, novel attacks are devised on the existing vulnerabilities.

The second challenge in this scope stems from the distributed nature of anomaly detection. While DL-based IDSes have proved themselves to be accurate, there is still the need to conform them to a distributed architecture from two practical points of view:

\begin{enumerate}
	\item It has been well discussed that relying solely on a single instance or sensor of an IDS will often yield inaccurate intrusion detection  \cite{bhargavi2013semantic}. Large and complex network architectures will require an ensemble of IDSes, each strategically placed in a specific location, ensuring optimal security and robustness  \cite{iyengar2020evaluation,seresht2014mais}. Furthermore, the collective knowledge of these scattered IDSes can be shared with a central unit to produce more comprehensive information and awareness regarding the network  \cite{chai2020fedat}.
	
	\item While relying on DL models, handling concurrent flows is not trivial. In most large networks, online traffic consists of many concurrent and interleaving flows. Each flow has a different start, end, and duration time. Consequently, considering a specific time window, the traffic consists of packets belonging to different flows. On the other hand, DL models need the sequence of a particular flow's packets to determine the flow label. As a result, these interleaving packets cannot be fed into a single DL model, and the flows should be separated beforehand.
		
\end{enumerate}

\comment{These points (\ie, cooperative distributed intrusion detection and interleaving packets)imply that it would be more advantageous to design IDSes in a distributed manner. While DL-based IDSes have proved themselves to be accurate, there is still the need to conform them to a distributed architecture from a practical point of view.}

The third and last challenge is that the performance of an online IDS depends on its ability to determine the correct flow label by inspecting fewer number of packets (\ie, early attack detection). A reliable and fast attack detection can stop the attack earlier and mitigate its full impact on the target organization.
Similar to the applicable traditional online IDSes, the aim is to determine the flow's label with some reliability per each packet arrival. When the IDS analyzes more flow packets, it increases its reliability score of the flow label. Security administrators can determine the thresholds of acceptable reliability scores according to the sensibility of the organization's assets. 

To summarize, the contributions of this paper are as follows:
\begin{enumerate}
	\item We design a novel framework for DL-based online IDSes that simultaneously addresses the three practical weaknesses that these systems are currently facing: the continuous adaptation to the network concept drift, early attack detection (\ie, determining the probability of a flow label by observing each incoming packet), and functioning efficiently in a multi-agent (\eg, multi-sensor) environment.

	\item We conduct extensive experiments and analysis to demonstrate the effectiveness of the proposed framework from different perspectives. We show that by exploiting deep continual learning methods, the proposed framework can adapt the IDS to new patterns in the network with a relatively small number of new flows (\ie, 128). Additionally, by utilizing LSTM models, the proposed framework is able to detect the intrusions of the state-of-the-art datasets CIC-IDS2017 and CSE-CIC-IDS2018 by just observing their first 15 packets. Furthermore, we show that the proposed framework performs well in a multi-agent environment, and different IDSes are able to effectively share their obtained attack knowledge, resulting in more reliable and robust intrusion detection.
\end{enumerate}

The rest of this paper is organized as follows. In the next section, we first review the related works in DL-based intrusion detectors, deep continual learning methods, packet labeling, and deep federated learning. In Section \ref{sec:framework}, we describe the proposed framework for online intrusion detectors.
Section \ref{sec:evaluation} presents details of experiments, dataset preprocessing, and evaluation results of the framework implementations.
Section \ref{sec:discussion} discusses and analyzes the results of the experiments and explores some possible future directions. Finally, Section \ref{sec:conclusion} concludes the paper.
	
	\section{Related Works}\label{sec:related}
	\subsection{Deep Learning-Based Intrusion Detection}
	Due to the capabilities of deep learning algorithms, including auto-extraction of suitable features, processing high dimensional data (e.g., content bytes of a flow), and supporting the time-series nature of the data, many studies have applied them in the scope of network intrusion detection. In the following, we review some of these research studies.
	
	In \cite{yin2017deep}, the authors employ recurrent neural networks (RNN) for intrusion detection and evaluate the performance for both binary and multiclass classification over the NSL-KDD dataset. In \cite{vinayakumar2017applying}, the intrusion detection application of different architectures of CNN-based DL models (\eg, CNN, CNN-RNN, CNN-LSTM, and CNN-GRU) are evaluated using the KDDCup 99 dataset. In \cite{ilango2022feedforward}, feedforward–convolutional neural networks (FFCNN) are exploited to detect low-rate denial service attacks in the internet of things (IoT). They utilize support vector machines (SVM) to indicate the important features used for detection iteratively, and their proposed scheme is evaluated on the CIC DoS 2017 dataset.
	
	The authors of  \cite{thakur2021intrusion} propose a Generic-Specific auto-encoder (GSAE) based model for considering the different distributions of attack domains. Their model concatenates the results of a generic auto-encoder and some domain-specific encoders as the input of a random forest (RF) model. They analyze their model on the CIC-IDS2017 dataset by evaluating web and DOS attack domains. 
	
	Distributed intrusion detection using mobile agents is discussed in \cite{riyad2019adaptive}. Each mobile agent analyses the traffic and detects the threats independently. Consequently, this distribution operation evades the single point of failure problem. Additionally, they propose algorithms for reducing false positives by using inter-agent communications. They use the principal component analysis (PCA) algorithm to select the traffic features. Then, an ensemble of support vector machines (SVM), artificial neural networks (ANN), and RF algorithms classify the input traffic. The evaluation has been done on the KDD99 dataset.

	
	Employing reinforcement learning (RL), particularly deep Q-learning, in network intrusion detection systems is the main contribution of the proposed framework in  \cite{kim2019designing}. The authors use two deep auto-encoder in their RL framework. One is for training the Q-learning model, and the other is for updating the model. The framework periodically applies mini-batch updates or Q-learning updates to make the model more adaptable to the continual evolution of cyber-attacks.
	
	A deep learning self-adaptive approach is presented in  \cite{papamart2019}. This approach consists of a transformation layer (the encoder) and a supervised learning deep model. It depends highly on the \emph{change signals} from the network mapper modules. Such entities should determine any network changes such as running services, available hosts, the operating system, and potential vulnerabilities. The approach learns a new auto-encoder model based on the stored traffic related to the signal period time and an archived initial labeled dataset. Then, it uses the encoder part as the new transformation layer by receiving the change signal. As a result, the model adapts itself to the new traffic distributions.
	A weakness of the mentioned approach is that in many cases, receiving change signals from a network mapper is not a reasonable assumption for changing the model. For example, sensing a change in the network load may result from a DDoS/DoS attack. More generally, the model should not adapt its transformation layer according to the change signals triggered by attacks. 

CSE-IDS  \cite{GUPTA2022102499} focuses on the imbalanced nature of classes in the network security scope. It proposes a three-layer deep learning-based IDS and assumes three traffic categories: benign traffic, majority attacks with frequent samples, and minority attacks that represent infrequent ones. A cost-sensitive deep neural network (DNN) separates the benign traffic from the malicious ones in the first layer. The cost-sensitive loss function handles the imbalanced number of attacks and benign traffic. Then, a boosting ensemble, namely eXtreme Gradient Boosting,  separates the suspicious samples into the benign class, different majority attack classes, and a single class representing all minority classes. Finally, an RF classifies the minority attacks into their respective classes. Besides, layer 2 and layer 3 use two oversampling techniques, namely, random oversampling and SVM-SMOTE. Their evaluation is based on the pre-extracted features of the NSL-KDD, CIDDS-001, and CIC-IDS2017 datasets.

In  \cite{WANG2021102177}, the authors integrate the stacked denoising auto-encoder (SDAE) (for reducing the noise of network traffic) and the extreme learning machine (ELM) (for increasing the IDS speed) as the SDAE-ELM model. This model is presented for a network intrusion detection system (NIDS). Besides, they propose to integrate the deep belief networks (DBN) (for extracting features from the log files of each host) and the softmax classifier (for determining the attack types) as the DBN-Softmax for the host-based intrusion detection system (HIDS). Their models use unsupervised data for the pretraining phase (learning the DAE and DBN layers of the NIDS and HIDS, respectively). Then the fine-tuning phase uses supervised learning for training the SDAE-ELM and DBN-Softmax. The authors evaluate the NIDS based on the pre-extracted features of KDD99, NSL-KDD, UNSW-NB15, and CIDDS-001 datasets. Additionally, the AFDA-LD dataset is used to evaluate the HIDS model.

The authors of  \cite{WANG2022102542} evaluate the ability of different machine learning-based models to detect encrypted malicious traffic. For the evaluation, they use the pre-extracted features of the datasets that contain malicious encrypted traffic. The candidates are the UNSW NS2019 and different versions of the CIC datasets, including ISCX IDS 2012, CIC-IDS2017, and CIC-ANDMAL2017.

Cretu-Ciocarlie et al. \cite{Cretu2009anomaly} propose an ensemble of n-gram based anomaly detectors (\ie, micro-models). The voting scheme determines the predicted label of the evaluation traffic. They use time-delimited slices of the dataset for training the disjoint micro-models. Additionally, the model updates itself by generating new models according to the recently received traffic. The new micro-models take the place of the oldest ones. Accordingly, the intrusion detector can be adaptable to the traffic concept drift.

Another chunk-based learning scheme is presented in  \cite{FOLINO202148}. The authors use disjoint time-delimited chunks of the training dataset for training a series of DNN classifiers. In addition, they use a combination of skip-connections, dropout, and cost-sensitive loss to manage the imbalanced training data.

In  \cite{soltani2021adaptable}, the authors propose DOC++ as a deep novelty-based classifier to detect not seen traffic (both the zero-day attacks and new benign behaviors). Besides, using a joint deep clustering algorithm, enough pieces of each new novel class evidence are gathered and used in the supervised labeling process and corresponding updating phase. The update process that is responsible for learning the new labeled concepts uses an active-passive strategy as the following steps:
\begin{enumerate}
\item Clone the existing active model to a passive model.
\item Run the cloned model's training, clustering, and post-training phases.
\item Migrate the traffic to the new model.
\end{enumerate}

\comment{Even though many research studies like  \cite{alrawashdeh2016toward},  \cite{kim2019designing}, and some of the studies mentioned above use}

Even though the above-mentioned studies and many other similar research use terms like deep learning-based online/real-time NIDS, most of them solely focus on improving the detection speed and accuracy (\ie, detection rate) of their models in comparison with the other approaches. Speed and accuracy are critical parameters in a real-world NIDS, but there are many other practical challenges in online NIDSes. For example, packet interleaving is an issue in real network traffic: packets of different flows are interleaved, and the proposed system should consider this challenge. 
Furthermore, network traffic concept drift is a prevalent phenomenon, and a practical IDS should adapt itself to these continuous changes. Additionally, a practical NIDS should determine the flow label upon receiving each packet and declare a reliability score for its decision. Measuring the performance of an online IDS is based on its capability to determine the true flow label with acceptable reliability by observing fewer packets of a flow. 

However, to the best of our knowledge, the above-mentioned challenges have not been investigated in most of the research studies related to online deep learning-based NIDSes.


\subsection{Sequence Labeling}
As mentioned before, an ideal characteristic that an IDS should possess is the ability to determine whether a flow is categorized as a possible threat in a gradual manner. 

To be more precise, since the packets corresponding to a flow do not arrive simultaneously, with the arrival of the first packet of a flow, the IDS presents an initial probability regarding the possibility of whether that flow is an attack. As time progresses, with the emergence of further packets, the IDS should produce a more accurate likelihood regarding that flow. 

One should bear in mind that in conjunction with conforming more to real-world scenarios, this scheme tends to be more efficient since there will be no need to allocate time and computational resources to accumulate all packets of a flow  \cite{hwang2019lstm}.
For this purpose, the IDS needs to be capable of handling two essential tasks:
\begin{enumerate}  
\item  Producing labels for each packet individually, rather than yielding a single label for the flow.
\item Using temporal features for estimating the probability. In other words, the IDS should also consider the previous packets of a flow in the inference process for a new packet.
\end{enumerate}  
Due to their ability to preserve memory over sequential inputs, RNN networks, LSTMs specifically, have been widely exploited in several domains (since they excel in circumventing the vanishing gradient problem  \cite{hochreiter1997long}). For instance, in the field of natural language processing (NLP), the research studies (\eg, \cite{ma2016end} and  \cite{huang2015bidirectional}) have used LSTMs to tackle sequence labeling tasks like \emph{part of speech tagging} and \emph{chunking}.

Similarly, some researchers have utilized LSTMs for network traffic classification. In  \cite{hwang2019lstm}, network traffic classification is done at the packet level by mapping this task to a sentence classification problem in NLP. This approach considers packets and their headers as sentences and words, respectively. The headers of a packet are used to construct a 64-dimensional word vector, and that vector is used as the input for an LSTM model to perform the classification. 
In   \cite{lopez2017network}, although several networks comprising LSTM segments have been designed to classify packets sequentially, they require the entire flow for classification. In \cite{mirza2018computer}, the authors employ different LSTM-based Auto-encoders to perform intrusion detection. To be more precise, a threshold is assigned to the reconstruction error of the model on network data sequences for classifying them as normal/anomaly. 

In \cite{ansari2022gru}, the authors employ deep models with gated recurrent units (GRU) to generate alerts for malicious sources. In their approach, a model is trained to learn the dependencies between previously generated alerts and predict future alerts for a malicious source.

In \cite{gao2019lstm}, both a many-to-many and a many-to-one LSTM are designed to address intrusion detection systems for the supervisory control and data acquisition (SCADA) protocol, and their results are compared.

Since a many-to-many LSTM model can classify packets individually and sequentially, our approach utilizes this technique as one of the base DL models inside the proposed adaptive framework. Furthermore, LSTMs can work with variable length input sequences (\ie, flows) \cite{lee2021towards}, thus making them more efficient and practical.

\subsection{Deep Continual Learning}
\label{sec:clrel}
In the proposed framework, our primary attention has been devoted to a specific family of online learning algorithms named \textit{Continual learning (CL)}, defined as the ability to learn new tasks that arrive sequentially by efficiently exploiting the knowledge learned in previous tasks \cite{van2019three}. The main dilemma in CL is a phenomenon called \textit{catastrophic forgetting}, characterized by the model performing poorly on the old tasks when trained on the new ones.

In recent years, valuable methods have been proposed to mitigate the problem of catastrophic forgetting for continual learning. In the following, we will review the two main categories related to our research.

\subsubsection{Continual Learning Based on Regularization}
\label{related:continual-reg}
These techniques exploit different regularization terms and constraints to avoid detrimental weight changes when training on new tasks. One naive solution would be to use an L2-Regularization term, but this approach will prevent the model from efficiently learning new tasks.

A ground-breaking technique known as elastic weight consolidation (EWC) is proposed in \cite{kirkpatrick2017overcoming}, which uses a regularization term based on the diagonals of a set of Fisher information matrices to reduce the plasticity of the weights of greater importance to the previous tasks.
The values on the diagonal of the Fisher information matrix measure the amount of information that the training samples provide for each parameter (\ie, weight) of the trained DL model, thus representing an importance factor for each weight. To be more precise, based on the definitions in \cite{martens2020new, van2019three}, the $i_{th}$ element of the Fisher information matrix diagonal is proportional to the expected value (\ie, based on the training data distribution) regarding the Hessian of the model's output with respect to the $i_{th}$ weight. Consequently, a high Hessian for a weight signifies the plasticity of the gradient of the model output based on that weight. Note that in a given task, the weights obtained from the training (\ie, optimization) procedure often represent a local minimum for the desired loss function. As a result, changes made to parameters with a high hessian would result in a substantial drift from that minima, resulting in a performance decline of the model on the mentioned task.

Since in EWC, the number of quadratic terms would increase linearly with the advent of new tasks, online EWC is proposed in  \cite{schwarz2018progress} which uses a single Fisher information matrix and updates it each time it learns a new task. Another method named synaptic intelligence (SI) is proposed in  \cite{zenke2017continual}. Instead of the Fisher information matrix, it tries to compute an online importance factor for each weight which describes its importance across all previously learned tasks.

\subsubsection{Continual Learning Based on Expansion} A number of approaches focus on the main idea to expand the network capacity by adding new layers or extending the previous layers to accommodate the knowledge associated with the new task \cite{rusu2016progressive}\cite{yoon2017lifelong}\cite{jain20213d_den}.

Progressive neural networks (PNN), as described in \cite{rusu2016progressive}, are models comprised of columns that each preserve a connection with all of their predecessors. Each column can be considered an individual network with a fixed architecture that includes blocks representing a network layer. A new column is added to the model with the arrival of new data, and training is done via freezing the previous columns. The main drawback of this approach is the constant, substantial increase of the network size for every new task, thus making it infeasible to maintain in the long run. Several methods have been proposed to circumvent this flaw by expanding the network as efficiently as possible.

As described in  \cite{yoon2017lifelong}, dynamically expandable networks (DEN) try to design an architecture that dynamically increases the network capacity when faced with new training data. At its core, a DEN first aims to modify the current network to perform well on the new data. In case of failure, each layer will be augmented by adding a fixed number of nodes, and the whole expanded network will be trained on the new data with the group sparse regularization \cite{scardapane2017group}. Due to this regularization term, some added nodes will be considered redundant after training and be pruned, thus preventing the network from becoming too large. In the end, if the weights of some previous nodes experience significant alteration during training, a duplicate of those nodes will be added to the network, and the network will be trained again.

One recent variation of DEN named 3d\_DEN is proposed in  \cite{jain20213d_den} for continual multi-class classification. Each task represents a new class, and a corresponding output node will be added to the network. In this approach, when training the network after expansion, only the added segments are trained, and the previous parts of the network are frozen, thus protecting them from catastrophic forgetting.

Since DEN and its variations rely on multiple sparse regularization terms, the high number of hyperparameters will make tuning the ideal network arduous. For this means, reinforced continual learning (RCL) is introduced in  \cite{xu2018reinforced}. In this method, for expanding the network, an LSTM network is used via reinforcement learning and policy gradient to predict the optimal number of nodes that should be added to each layer, with respect to both the detection rate and size of the network.

Although the approaches mentioned above try to expand the network as efficiently as possible, the network's size will still grow after each task, which is considered a drawback in the long run. An approach for fully compressing the network after the expansion is proposed in regularize, expand, and compress (REC) framework \cite{zhang2020regularize}. Similar to RCL, REC exploits reinforcement learning (AutoML  \cite{sutton2000policy}) to expand the network. The whole network is trained on the new task with regularizations based on multi-task learning and the Fisher information matrix. After that, the compression is done using the knowledge distillation approach \cite{hinton2015distilling} and soft labels; thus, the network reshapes to its original architecture.

\subsection{Deep Federated Learning}
Federated learning (FL) is an ML approach for training a model by utilizing distributed devices that contribute to the training process based on their local data. Both synchronous and asynchronous methods have been proposed to this end, but since the nature of our problem requires an asynchronous setting, we will mainly focus on the latter.

In  \cite{gimpel2010distributed}, an asynchronous distributed optimization algorithm is designed, which despite a minor error in the training procedure, performs well when evaluated on NLP tasks. In  \cite{xie2019asynchronous}, an asynchronous federated learning scheme is proposed in which each worker independently trains a model with a regularization term that prevents any significant drift from the main model. Also, the main model is updated via weighted averaging with the worker model.
 
In  \cite{chai2020fedat}, a federated learning system is designed based on dividing the clients into different groups called tiers. In this approach, a tiering module partitions the clients into tiers based on their performance (e.g., response latency). In each tier, the updating process is synchronously performed by the tier members via gradient computation and optimization. Furthermore, The main model gets updated asynchronously based on the weighted averaging of the models obtained from the tiers. 

In \cite{diro2018distributed}, the authors propose a distributed attack detection mechanism for IoT based on fog computing \cite{yi2015survey}. In their approach, the fog nodes are responsible for locally training DL models that act as intrusion detectors at the network edge. Furthermore, a coordinator master is used to propagate the local updates and parameters between the fog nodes, and this optimization procedure is conducted via distributed SGD.

These proposed schemes and designs provide a solid foundation for designing our multi-agent framework as described in Section \ref{sec:framework}.

	\section{Framework}\label{sec:framework}
	
	In this section, our proposed online anomaly-based intrusion detector is described. There are three challenges for an applicable intrusion detection system. The first challenge relates to the emergence of new attacks and benign user/traffic behavior changing over time. To address this continuous adaption challenge, we use deep continual learning methods, as we will discuss later in this section. 
	
	As mentioned before, another challenge of an online IDS is making a progressive decision about a flow by observing the stream of its packets. The reason is that the best online IDSes are the ones that can detect an attack with fewer packets. In other words, threat detection should be done before the attacker completes the attack.
	
	The third challenge of an online IDS corresponds to the interleaving nature of the packets of different flows in the network traffic. In particular, to address this issue and consider a high throughput network, we propose to use a distributed architecture for handling all packets of each flow in an agent. In this architecture, each agent implements a sub-model of the main DL model. Then, to update the model, the distributed sub-models are aggregated in the main deep anomaly-based model.
	
	The above-mentioned challenges are discussed in the following sections.
	
	\subsection{Continuous Adaption to Network Concept Drift}
	\label{sec:continual-leraning}
	
	With the advent of a new attack, we expect our IDS to conform itself to the new data, and while preserving its ability to detect the previously learned abnormalities, it should extend its knowledge to recognize the new one. 
	To achieve this goal, we employ a continual learning-based algorithm that best satisfies the needs and constraints of an IDS.
	
	The proposed algorithm assumes a DL model consists of two parts: the base part and the dense part. The base part usually consists of either LSTM or convolutional (CNN) layers and is followed by the dense part that comprises multiple fully-connected (FC) layers.
	Ensuing from the deductions made in  \cite{jain20213d_den} and  \cite{yosinski2014transferable}, the base parts serve as a pre-trained and frozen section of our network, whereas the FC layers will change and train continually on new anomalies. This approach has two main benefits:
	\begin{enumerate}
		\item The base part will determine the general features of our inputs (may it be flows or individual packets, as described in Section \ref{sec:evaluation}), and the FC layers will both learn new specific features and better classify the general features by training on new data.
		\item Each continual training will require less computation since the pre-trained network will not be involved.
	\end{enumerate}
	
	The proposed continual learning algorithm, similar to those mentioned in Section \ref{sec:clrel}, is based on the expansion approach, \ie,
	each FC layer is augmented with a set of nodes. More specifically, each added node will have inputs from all nodes in the previous layer (including the augmented ones), but its outputs will only be connected to the new nodes in the next layer, thus allowing it to capture new features while not altering the nodes from older tasks (i.e., attacks in the security scope)  \cite{jain20213d_den}. In this expansion phase, based on prior work, we are provided with two options:
	\begin{itemize}
		\item Adding a fixed number of nodes to each layer (denoted as $k$)
		\cite{yoon2017lifelong, jain20213d_den}.
		\item Designing a controller for configuring the optimal numbers of nodes for each layer based on RL approaches (which have been used prevalently in the network anomaly detection scope \cite{adawadkar2022cyber}). To be more precise, each time the controller generates the number of nodes corresponding to each layer, it receives a reward and updates itself via policy gradient techniques. This process is repeated several times until the best result is achieved  \cite{xu2018reinforced, zhang2020regularize}.
	\end{itemize}
	Although the latter approach tends to discover a more efficient expanded network, our analysis indicated that the former would be better suited to our domain, as explained below.
	
	First, the latter approach will require a substantial amount of time to find the optimal network, which is a major flaw since the IDS is expected to perform on a real-time basis. Each time the controller predicts the number of added nodes, training has to be conducted on the corresponding child network to yield a reward for the controller. This process might have to be carried out several times to yield the best result. On the other hand, using a fixed number of nodes will require training the expanded network only once.
	
	Second, since the expansion procedure is ensued by compression (as described in Section \ref{sec:framework_federated}), finding the minimum number of nodes in each layer is not necessary. Also, in contrast to  \cite{yoon2017lifelong} and \cite{jain20213d_den}, there is no need to perform $l_{1}$-norm, or group sparsity regularization, and tuning their corresponding hyperparameters for training on the new task, since compressing the network will not rely on this technique as explained in the next section.
	
	After adding $k$ nodes to each FC layer, the training phase consists of two sections:
	\begin{enumerate}
		\item The nodes pertaining to the previous tasks are frozen, and while only the added nodes are kept trainable, training is done on the data of the new task (\ie, new attack). As mentioned above, there will be no need for  any kind of regularization. Hence, the training can be described as optimizing a single loss function, \ie, 
		\begin{equation} \label{eq: train1}
			\begin{aligned}
				\min_{W^\mathsf{Add}} \mathcal{L} \Big( W^\mathsf{Add} \big| W^\mathsf{Prv} , \mathcal{D}_{\mathsf{train}} \Big),
			\end{aligned}
		\end{equation}
		where $\mathcal{L}$ is our desired loss function (\eg, binary cross-entropy), $W^\mathsf{Add}$ describes the weights of the newly added nodes to the network, $W^\mathsf{Prv}$ represents the (frozen) weights of previous nodes, and $\mathcal{D}_{\mathsf{train}}$ is the dataset comprising the new traffic for training.
		
		\item After the first step, the expanded model's performance is measured on a validation set $\mathcal{D}_{\mathsf{val}}$, and in the case its detection rate is below a preset threshold $\tau$, instead of solely training the added nodes, the whole network is trained under the following equation
		 (\cite{zhang2020regularize})
		 
			\begin{equation} \label{eq: train2}
				\min_{W^\mathsf{Exp}}\left[
				\begin{aligned}
					\mathcal{L}(W^\mathsf{Exp} |  \mathcal{D}_{\mathsf{train}})  +\lambda_{1}\sum_{i = 1}^{\mathcal{N}_\mathsf{params}}\mathcal{F}_{ii}^\mathsf{Prv}(\theta_{i}^\mathsf{Exp} - \theta_{i}^\mathsf{Prv}) +  \lambda_{2}\norm{[W^\mathsf{Exp} ; W^\mathsf{Prv}]}_{2,1} + \lambda_{3}\norm{W^{\mathsf{Add}}}_{1}
				\end{aligned}
			\right],
			\end{equation}
		
		where $\mathcal{N}_\mathsf{params}$ is the number of weights in the model prior to expansion, $W^\mathsf{Prv} = \{\theta_{i}^{\mathsf{Prv}}\}_{i=1}^{\mathcal{N}_\mathsf{params}}$, as introduced above, are the weights of the model before expansion, $W^{\mathsf{Add}}$ are the weights of the newly added nodes, and $W^\mathsf{Exp}$ are all the weights of the expanded model. In the expanded model, using the Fisher information matrix diagonal, the term $\sum_{i = 1}^{\mathcal{N}_\mathsf{params}}\mathcal{F}_{ii}^\mathsf{Prv}(\theta_{i}^\mathsf{Exp} - \theta_{i}^\mathsf{Prv})$, is enforced on the weights corresponding to the previous task to avoid catastrophic forgetting (as discussed in Setion \ref{related:continual-reg}). The term $\norm{[W^\mathsf{Exp} ; W^\mathsf{Prv}]}$ is an $l_{2,1}$-norm regularization \cite{zhang2020regularize} (\ie, $\norm{\norm{W^\mathsf{Exp}}_{2}, \norm{W^\mathsf{Prv}}_{2}}_{1}$) term derived from multi-task learning, aiming to learn the shared representations between the weights of the model prior to and after expansion, and $\norm{W^{\mathsf{Add}}}_{1}$ is a sparsity-inducing regularization term   \cite{gong2012multi} imposed solely on the new nodes for efficient learning of the features specific to the new traffic.
		
		Furthermore, for practically computing the diagonal of the Fisher information matrix, we employ the method proposed in \cite{van2019three}. Namely, for the $i_{th}$ element of the diagonal we have:
		\begin{equation} \label{eq: Fisher}
			\begin{aligned}				
				\mathcal{F}_{ii} = \frac{1}{|S|}
				\sum_{(x,y) \in S} \frac{\delta \log p(Y = y|x, \theta)}{\delta\theta_{i}},
			\end{aligned}
		\end{equation}
		
		where $\mathcal{F}_{ii}$ is the $i_{th}$ element of the Fisher information matrix diagonal (\ie, corresponding to the $i_{th}$ weight of the model) and $S$ is the data set used for training the model. Furthermore, $\theta$ are the weights of the model after training, $(x,y)$ represents any labeled sample from $S$, and $p(Y = y|x, \theta)$ is the produced probability by the model for the correct class label. Moreover, a proposed strategy is discussed for updating the Fisher information diagonal throughout the continual learning procedure in Section \ref{sec:framework_federated}.
	\end{enumerate}

	Algorithm \ref{alg:con} describes the proposed continual learning procedure.
	
	
	\begin{algorithm}
		\caption{Continual Learning Algorithm.}\label{alg:con}
		\begin{flushleft}
			\textbf{Input :} \\
			\hspace*{\algorithmicindent} \emph{$\mathcal{D}_\mathsf{train}$} : New dataset to train on \\
			\hspace*{\algorithmicindent}
			\emph{$\mathcal{D}_\mathsf{val}$} : Validation dataset \\
			\textbf{Output : } \\
			\hspace*{\algorithmicindent}$W^\mathsf{Exp}$ : The weights of the expanded network
		\end{flushleft}
		
		\begin{algorithmic}[1]
			\State Add $k$ units to all layers
			\State Obtain $W^\mathsf{Exp}$ by training the network based on \eqref{eq: train1}
			\If{ detection rate of $W^\mathsf{Exp}$ model on $\mathcal{D}_\mathsf{val} <  \tau$}
			\State Obtain $W^\mathsf{Exp}$ by training the network based on \eqref{eq: train2}
			\EndIf
		\end{algorithmic}
	\end{algorithm}

	\subsubsection{Data Sampling}
	In some cases, incrementally training solely on a new set of data samples from unknown traffic might make our model biased towards that new traffic, which will be an instance of catastrophic forgetting. As suggested in  \cite{jain20213d_den,soltani2021adaptable}, with the advent of new data, we will constitute a training set that possesses the new data in conjuncture with samples corresponding to the previous attacks and benign flows that the model has been previously trained on.
	
	To implement this approach, the collective number of data samples belonging to the previous attacks should be equal to the number of the new attack samples. Since our model is a binary classifier between benign and attack flows, the number of benign samples should be equal to the total number of attacks (\ie, including the old and new attacks).
	Algorithm \ref{alg:dsa} describes this procedure in detail. 
	
	One should bear in mind that sustaining all the previous instances is evidently unfeasible for practical scenarios. However, as discussed in Section \ref{sec:discussion}, the proposed updating strategy is able to adapt the model to new traffic with a small number of instances. Consequently, it suffices to preserve a limited number of instances from previous flows to prevent bias (\ie, set a threshold for the maximum number of previous benign/attack samples). Furthermore, another practical approach for reproducing previous samples would be using Generative Adversarial Networks (GAN) that can support continuous updating to new data \cite{seff2017continual, liang2018generative, varshney2021cam, andresini2021gan}.
	
	\begin{algorithm}
		\caption{Data Sampling Algorithm.}\label{alg:dsa}
		\begin{flushleft}
			\textbf{Input :} \\
			\hspace*{\algorithmicindent} \emph{$\mathcal{R}$} : raw samples of the new traffic\\
			\textbf{Output : } \\
			\hspace*{\algorithmicindent}$\mathcal{D}_{t}$ : augmented dataset for the new traffic (\ie, new task)
		\end{flushleft}
		\begin{algorithmic}[1]
			\State $\mathcal{B}$ = dataset containing benign samples
			\State $\mathcal{A} = (\mathcal{A}_1,\mathcal{A}_2,\ldots,\mathcal{A}_{t-1})$: datasets of previous attacks
			\State $\mathcal{D}_{t} = \mathcal{R}$
			\State Split $\mathcal{R}$ to $\mathcal{A}_{t}$ (new attack samples) and $\mathcal{B}_t$ (new benign samples)
			\State $s_A =  \mathrm{len}(\mathcal{A}_t)$
			\For{\texttt{$i = 1,2,\ldots,t-1$}}
			\State Choose $s_A$ samples from $\mathcal{A}_i$ and add to $\mathcal{D}_{t}$
			\EndFor
			\State $s_B = \mathrm{len}(\mathcal{D}_{t}) - \mathrm{len}(\mathcal{B}_t)$
			
			\State Choose $s_B$ samples from $\mathcal{B}$ and add to $\mathcal{D}_{t}$
		\end{algorithmic}
	\end{algorithm}
	
	
	\begin{figure}[!h]
		\centering
		\includegraphics[width=0.8\linewidth]{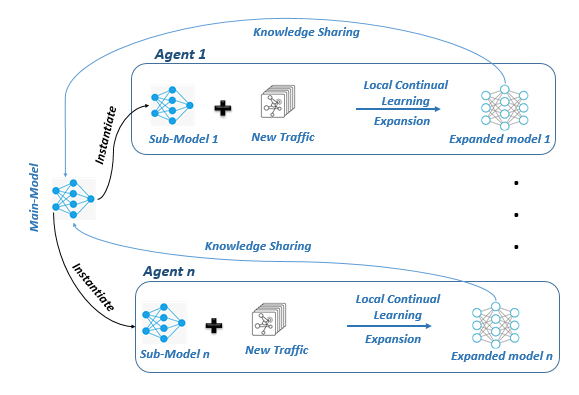}
		\caption{Overview of the proposed multi-agent Architecture. $Agent_{1},..., Agent_{n}$ (\eg,different IDSes) contain each a sub-model which is initialized with the the weights of the main-model. After learning a new anomaly based on Algorithm \ref{alg:con}, the main-model is updated via knowledge sharing (\ie, federated distillation).}
		\label{fig:1}
	\end{figure}
	
	\label{sec:data-sampling}

	\subsection{Multi-Agent IDS}\label{sec:framework_federated}
	
	To address the distributed requirements of an IDS (as discussed in Section \ref{intro}), we have proposed to employ a multi-agent federated learning architecture. Each agent is assigned a part of the traffic flows and captures the new abnormalities and benign traffic concept drift based on the assigned traffic, and then, updates itself.
	
	To be more precise, each agent consists of a \textit{sub-model} that continually learns new traffic behavior. Once an agent has finished its continual learning procedure, it asynchronously updates the \textit{main-model} through knowledge distillation \cite{hinton2015distilling}. Thus, the collective knowledge obtained and shared by all the agents will be incrementally integrated into the main-model.
	
	An overview of the proposed federated learning architecture is shown in Figure \ref{fig:1}.
	Each agent initializes the weights of its sub-model with the latest weights of the main-model prior to its continual learning procedure. After the learning phase, in order to update the main-model, each agent engages in an asynchronous optimization with the loss function using a combination of the logits (\ie, the input vector of the final softmax layer as the soft labels) and the actual labels (\ie, hard labels). In addition, in order to prevent catastrophic forgetting, a regularization term based on the diagonal of the Fisher information matrix of the main-model is exploited. Thus, in order to update the main-model through knowledge distillation, we propose the $\ell$th agent computes and sends to the main-model the gradients of the following loss function
	\begin{equation} \label{eq: distill}
		\begin{aligned}
			f_{\mathsf{dist}}(W_\mathsf{main}) = \mathcal{L}(W_\mathsf{main} ;  \mathcal{D}_\ell) + \mathcal{L}_\mathsf{kd}(W_\mathsf{main} ; \mathcal{Z}_\ell)  + \lambda\sum_{i = 1}^{\mathcal{N}_\mathsf{params}}\mathcal{F}_{ii}(\theta_\mathsf{main}^{i} - \theta_\mathsf{init}^{i}),
		\end{aligned}
	\end{equation}
	where again, $\mathcal{N}_\mathsf{params}$ is the total number of parameters in the main-model,  $W_\mathsf{main} = \{\theta^i_\mathsf{main}\}_{i=1}^{\mathcal{N}_\mathsf{params}}$ is the new weights of the main-model, $W_\mathsf{init}=\{\theta^i_\mathsf{init}\}_{i=1}^{\mathcal{N}_\mathsf{params}}$ and $\mathcal{F}$ are the weights and the Fisher information diagonal of the main-model prior to distillation,  $\mathcal{D}_\ell$ is the training data observed by the $\ell$th agent, and $\mathcal{Z}_\ell$ are the logits received through the expanded model.
	
	In order to asynchronously update the main-model, an agent first acquires the latest version of $W_\mathsf{init}$ and $\mathcal{F}$ from the main-model. Then, the main-models' parameters are updated through the following update rule (\cite{gimpel2010distributed})
	\begin{equation} \label{eq: distributed}
		\begin{aligned}
			W'_\mathsf{main} = W_\mathsf{main} - \mu  \sum_{\ell \in M}\nabla_{\ell}(f_\mathsf{dist}(W_\mathsf{main})),
		\end{aligned}
	\end{equation}
	where $\nabla_{\ell}(f_\mathsf{dist}(W_\mathsf{main}))$ is the gradient of $f_\mathsf{dist}(W_\mathsf{main})$ computed by the $\ell_{th}$ agent on its own batch. Also, $M$ is the set of agents that have sent a gradient in the time interval between the last two updates. 
	
	Once an agent's federated distillation procedure comes to an end, it also computes the Fisher information matrix diagonal based on the latest version of $W_\mathsf{main}$ and its own data, using Equation \ref{eq: Fisher}. This matrix is sent to the main-model, updating the main Fisher information matrix diagonal based on the following equation
	\begin{equation} \label{eq:fisher-upd}
		\begin{aligned}
			\mathcal{F'}_\mathsf{main} =(1 - \alpha) \mathcal{F}_\mathsf{main} + \alpha 	\mathcal{F}_\mathsf{agent},	
		\end{aligned}
	\end{equation}
	where $\mathcal{F'}_\mathsf{main}$ is the new Fisher information matrix diagonal of the main-model, $\mathcal{F}_\mathsf{main}$ is the diagonal of the previous Fisher information matrix of the main-model, $\mathcal{F}_\mathsf{agent}$ is the Fisher information matrix diagonal sent by the agent, and $\alpha$ is an aggregation weight.
	
	Based on the proposed federated learning architecture, the procedure that an agent undertakes to update the main-model is described in Algorithm \ref{alg:fed-agent}. Note that the proposed approach has the practical benefit of not expanding the main-model; thus, the main-model will not grow infinitely and can be practically applied in the long term without needing additional memory. Furthermore, the federated distillation procedure also functions as a compression mechanism for the agents. As a result, an agent's expanded model can be replaced with the updated main-model at the end of this process.

	\begin{algorithm}
		\caption{Agent Learning Procedure.}\label{alg:fed-agent}
		\begin{flushleft}
			\textbf{Input :} \\
			\hspace*{\algorithmicindent} \emph{$\mathcal{R}$} : Flows pertaining to the new traffic 	
		\end{flushleft}
		\begin{algorithmic}[1]
			\State Obtain \emph{$\mathcal{D}_{t}$} by using \emph{$\mathcal{R}$} as input to Algorithm \ref{alg:dsa}.
			\State Split \emph{$\mathcal{D}_{t}$} to \emph{$\mathcal{D}_\mathsf{train}$} and \emph{$\mathcal{D}_\mathsf{val}$} for training and validation.
			\State Get \emph{$W_\mathsf{main}$} and \emph{$\mathcal{F}$} from the main-model.
			\State Obtain $W_\mathsf{Exp}$ from Algorithm \ref{alg:con} using \emph{$\mathcal{D}_\mathsf{train}$}, \emph{$\mathcal{D}_\mathsf{val}$}, {$W_\mathsf{main}$}, and \emph{$\mathcal{F}$}.
			\State Update \emph{$W_\mathsf{main}$} and \emph{$\mathcal{F}$} from the main-model (in case that other agents have updated the main-model).
			\State Set $W_\mathsf{init} = W_\mathsf{main}$ 
			\For{each training step}
			
			\State Get mini-batch and labels from \emph{$\mathcal{D}_\mathsf{train}$} and the logits from $W^\mathsf{Exp}$.
			\State Compute the gradient of \eqref{eq: distill} and send it to the main-model.
			\State Wait for the main-model to send $W_\mathsf{main}$
			\EndFor
			
			\State Compute \emph{$\mathcal{F}_\mathsf{agent}$} based on \emph{$\mathcal{D}_\mathsf{train}$} and sent to the main-model in order to compute \eqref{eq:fisher-upd}.
		\end{algorithmic}
	\end{algorithm}

\begin{figure}[!h]
		\centering
		\includegraphics[width=0.7\linewidth]{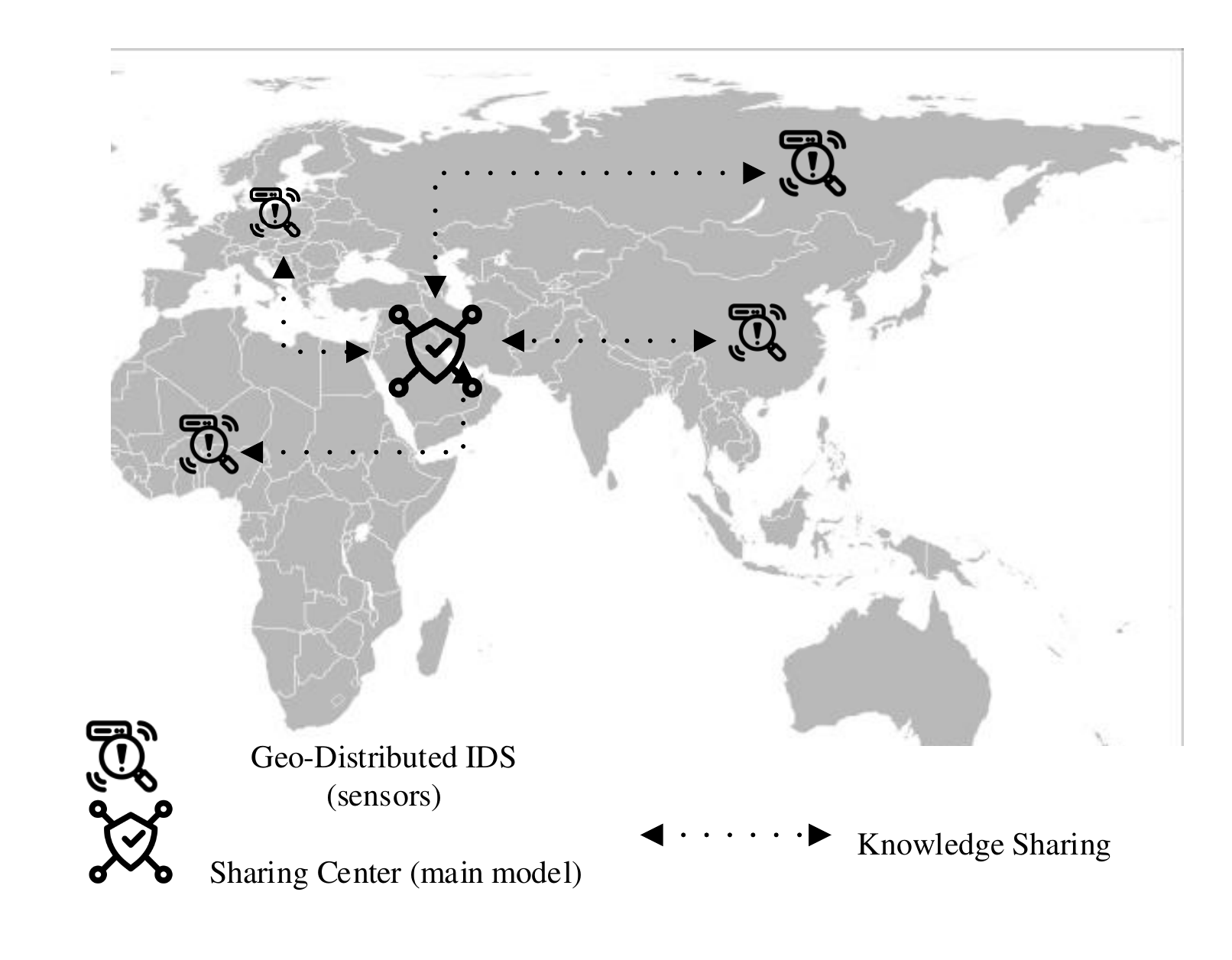}
		\caption{An illustration of the geo-distributed IDSes that can share their knowledge through a sharing center (\ie, main model).}
		\label{fig:geo-dist}
	\end{figure}

\begin{figure}[!h]
		\centering
		\includegraphics[width=0.7\linewidth]{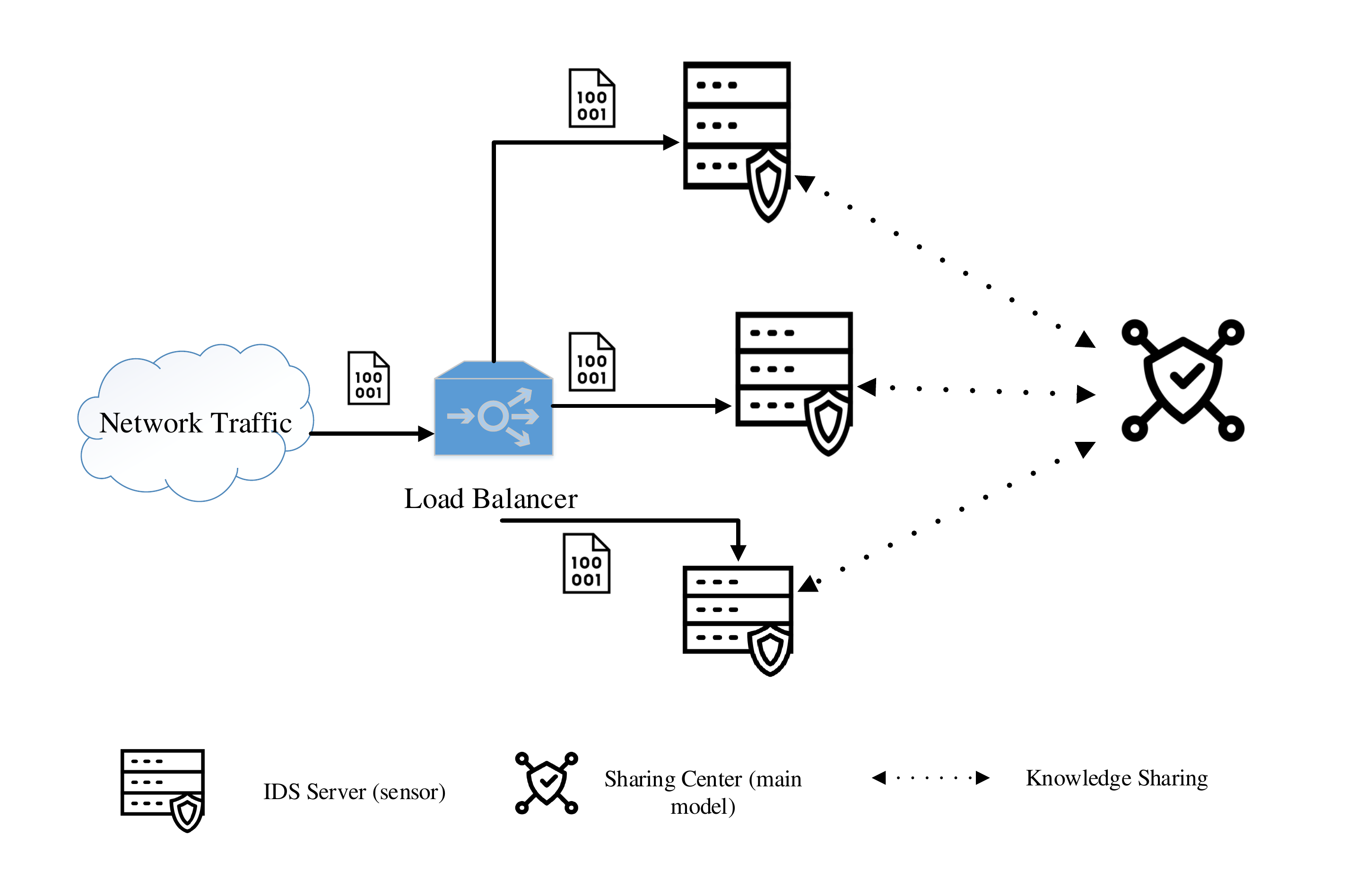}
		\caption{An illustration of load balancing in the proposed multi-agent framework.}
		\label{fig:load-balance}
		
	\end{figure}

	The proposed multi-agent architecture is advantageous in several aspects:
	\begin{enumerate}
		\item {In terms of privacy, since only gradients are exchanged between an agent and the main-model, the IDS can be shared between numerous organizations. Each can contribute to updating the main anomaly detection model while preserving their data privacy. Even on a geo-distributed scale, different IDSes and organizations scattered over various locations can all collaborate with the main model (\ie, sharing center) to securely adapt themself to new traffic patterns. An overall schematic of this scenario is depicted in Figure \ref{fig:geo-dist}.}
		\item {With the emergence of Big Data, IDSes have to face colossal and highly fast generated data streaming into the network \cite{othman2018intrusion}. A multi-agent architecture allows the dispersion of data among the sub-models in a parallel structure (\ie, load balancing the traffic flows, as demonstrated in Figure \ref{fig:load-balance}), improving the efficiency in both detecting intrusions and updating the IDS to new traffic behavior through a distributed training process.}
		\item{Interleaving traffic packets can be tackled by assigning each flow's packets to a specific agent (note that each agent can be assigned multiple flows).}
	\end{enumerate}
	
	\section{Experimental Evaluation}\label{sec:evaluation}
	\label{experiment}
	This section describes the evaluation details of the proposed framework to reproduce the experiments. First, the evaluation infrastructure, including preprocessing phase, evaluated datasets, and hardware specifications, are described. These are the common infrastructure for all the following experiments. Then, different deep online anomaly detectors' implementations are evaluated. The next step considers the online IDS challenge of progressively determining the flow label upon each packet arrival. Finally, the proposed distributed architecture for implementing a DL-based NIDS is evaluated.\footnote{The implementations of all evaluated models are available at \url{https://github.com/INL-Laboratory/Continual-Federated-IDS}.}
	
	\subsection{Evaluation Infrastructure}
	In this work, the deep intrusion detection (DID) framework introduced in \cite{soltani2020content} is used in the preprocessing phase of all experiments. The DID approach is selected for its ability to self-extract appropriate features and the capability of detecting a wide range of attacks, including content-based ones like SQL injection and Heartbleed attacks. The content-based attacks are the main segment of the threats with high malicious impacts on the targeted organizations. Consequently, this preprocessing phase can significantly affect the applicability of the proposed framework.
	
	As the DID approach is designed for the applicable IDSes, it requires the pure content of traffic flows (\eg, in PCAP format). Consequently, the scope of applicable datasets for evaluating deep IDSes is constrained to the ones that include the labeled content of the traffic. The privacy issues restrict the dataset developers from publishing the details of the real network traffic. As a result, datasets with entire traffic content such as DARPA 1999  \cite{lippmann20001999} (which is the base of the KDD99  \cite{KDD99} and NSL-KDD  \cite{tavallaee2009detailed} dataset), CIC-IDS2017  \cite{Sharafaldin2018ISCX}, and CSE-CIC-IDS2018  \cite{CIC2018} are all generated in an emulated network.
	
	In this work, to properly evaluate the proposed framework, we have used the more up-to-date datasets (CIC-IDS2017 and CSE-CIC-IDS2018), which have implemented the more recent attack types like SSH brute force botnet, DoS, DDoS, web, and infiltration attacks. Most importantly, they contain content-based attacks like SQL injection, XSS attacks, and Heartbleed. Additionally, benign profiles are extracted based on the abstract behavior of 25 users over the HTTP, HTTPS, FTP, SSH, and email protocols. Besides detecting the anomalies with a high detection rate, an IDS should produce low false-negative rates as well. As a result,
	in addition to anomaly flows, we use benign traffic in our experiments.
	
	In order to prepare the data to feed into the DL models, we use a packet size of 200 bytes and a flow size of 100 packets, resulting in a 20000-dimensional input vector (which we will refer to as the flow matrix). This selection is based on the analysis of the correspondent datasets investigated in  \cite{soltani2020content}. 
	To implement the proposed framework, we employ the Keras library  \cite{chollet2017} with Tensorflow  \cite{tensorflow2015-whitepaper} as its backend. The characteristics of our experimental environment are shown in Table~\ref{tab:Hardware}.
	
	\begin{table}[!h]
		\caption{The system specification of the experimental environment.}
		 \label{tab:Hardware}
		\centering
		\resizebox{0.5\linewidth}{!}{%
			\begin{tabular}{|c|l|}
				\hline
				\parbox[c][][c]{2cm}{\centering OS} & \parbox[c][0.9cm][c]{5cm}{Ubuntu Version 20.04.3 LTS with Kernel 5.4.0-81-generic
				}\\
				\hline
				CPU & \parbox[c][0.9cm][c]{5cm}{Intel(R) Core(TM) i7-6900K 3.20GHz
					with 16 virtual cores}\\
				\hline
				RAM & \parbox[c][0.5cm][c]{5cm}{32 GB}\\
				\hline
				GPU & \parbox[c][0.5cm][c]{5cm}{GeForce GTX 1080 Ti}\\
				\hline
				GPU Frame Buffer & \parbox[c][0.5cm][c]{5cm}{8 GB}\\
				\hline
			\end{tabular}
		}
	\end{table}
	
	\subsection{Model Architectures}\label{sec:architecture}
	We evaluate our proposed framework with two different architectures (i.e., CNN and LSTM). In the following, we describe each architecture's base and dense parts, as discussed in Section \ref{sec:continual-leraning}.
	
	In the first architecture (CNN-based), the base part comprises two consecutive  2D convolution layers with 8 and 16, a $3 \times 3$ kernel size, a stride of $1 \times 1$, and no padding. The dense part comprises four layers with 256, 128, 64, and 2 neurons, respectively. 
	
	The second architecture (LSTM-based) consists of a single, many-to-many LSTM layer with 1024 cells as the base part. Many-to-many LSTMs can generate separate outputs for each of the corresponding sequential inputs. The dense part has five layers with 512, 256, 128, 64, and 2 neurons. 
	
	The above-mentioned architectures use different input vectors. The first architecture uses the entire flow matrix as the input (\ie, the input is a matrix of size $200\times 100$). In contrast, the second architecture takes individual packets as the input (\ie, the input is a vector of size $200$, however, a sequence of $100$ such vectors are fed into the model) and estimates a probability for the flow label after processing each packet. Consequently, the second architecture is more applicable to early attack detection in IDSes.
	
	The ReLU activation function and a dropout of 0.2 are used in both architectures for all but the last layer. In the last layer of both architectures, a softmax function is implemented to compute the benign/anomaly probabilities.
	
	\subsection{Hyperparameter Settings}
	In order to obtain the best values of hyperparameters, including training batch size, epochs, and coefficients of the regularization terms, we employ a grid search procedure.
	Moreover, for the continual learning algorithm, we adhere to the method used in  \cite{jain20213d_den} for determining the number of added nodes (\ie, increasing the value of $k$ to the point where no improvement in the overall detection rate is witnessed).
	Table \ref{tab:Hyperparams} presents the chosen values of all hyperparameters used throughout the experiments, in addition to their searched space.
	
	\begin{table}[!h]
		\caption{The hyperparameters used in the evaluations.} \label{tab:Hyperparams}
		\centering
		\resizebox{0.6\columnwidth}{!}{%
		\begin{tabular}{|c|c|c|c|}
			\hline
			Parameter  & Usage    & Search Space &  Chosen Value        \\ \hline
			$\lambda_1$    & Equation \ref{eq: train2}         &\parbox[c][0.8cm][c]{3cm}{\centering  [$1^{-6}, 1^{-3}, 1 , 10$]}    & 1     \\ \hline
			$\lambda_2$   & Equation \ref{eq: train2}      &\parbox[c][0.8cm][c]{3cm}{\centering  [$1^{-6}, 1^{-3}, 1 , 10$]}       & $1^{-3}$ \\ \hline
			$\lambda_3$    & Equation \ref{eq: train2}     &\parbox[c][0.8cm][c]{3cm}{\centering  [$1^{-6}, 1^{-3}, 1 , 10$]}       & $1^{-3}$ \\ \hline
			$\lambda$     & Equation \ref{eq: distill}      &\parbox[c][0.8cm][c]{3cm}{\centering  [$1^{-6}, 1^{-3}, 1 , 10$]}         & 1     \\ \hline\hline
			batch size & Initial Training &\parbox[c][0.8cm][c]{3cm}{\centering  [8, 16, 32, 64, 128]}   & 32    \\ \hline
			epochs     & Initial Training &\parbox[c][0.8cm][c]{3cm}{\centering  [30, 50, 80]}   & 50    \\ \hline\hline
			batch size & Continual Learning  &\parbox[c][0.8cm][c]{3cm}{\centering  [8, 16, 32]} & 16    \\ \hline
			epochs     & Continual Learning &\parbox[c][0.8cm][c]{3cm}{\centering  [10, 20 , 30, 40]} & 20    \\ \hline
			$k$     & Continual Learning      &\parbox[c][0.8cm][c]{3cm}{\centering  [5, 10 , 12, 15]} & 10     \\ \hline\hline
			batch size & Federated Learning &\parbox[c][0.8cm][c]{3cm}{\centering  [8, 16, 32]} & 16    \\ \hline
			epochs     & Federated Learning &\parbox[c][0.8cm][c]{3cm}{\centering  [10, 20, 30 , 40]} & 20    \\ \hline
		\end{tabular}
	}
	\end{table}

	\subsection{Deep Adaptive Anomaly Detectors}
	\label{sec:DeepAdaptive}
	In this section, we devise two scenarios for evaluating the ability of models to learn new anomalies. Note that in the following experiments, we use the term \textit{known attack} for an attack class if a DL model has previously been adapted (\ie, trained or updated) to that attack. Furthermore, the \textit{zero-day attack} term is used for an attack class that the model has not been adapted to. In the first scenario, we use a pairwise evaluation: one known attack alongside one zero-day (\ie, new) attack. In the second scenario, we aim to evaluate a model's ability to learn consecutive new anomalies over time, \ie, some anomalies learned continually over time (as known attacks) and one zero-day attack.
	
	In the first scenario's experiments, a model is initially trained with a sufficient number of flows (\ie, about 3000\textasciitilde5000) from benign and one known attack (\ie, anomaly). Afterward, for each of the remaining attacks (as the new anomalies), a set of 128 flows are used to train the expanded initial model (see Algorithm \ref{alg:con}). Then, the expanded model is compressed back to its initial architecture (see Equation \ref{eq: distill}). Note that to resemble a more practical circumstance, in the above, the number of unknown attack flows is selected relatively small for evaluating the adaptive IDS.
	
	In the evaluation phase, we report the models' detection rate for known and unknown attacks according to two separate datasets created from the original data, \ie, known and zero-day datasets. The first one contains 500 known attack flows, and the second one includes 500 zero-day attack flows. Additionally, 500 benign flows are added to both datasets.
	
	The results of the above-mentioned scenario's experiments are shown in Tables \ref{CNN:cont-2017} and \ref{LSTM:cont-2017}. The first column indicates the experiment's known attack, which will be used to train the initial model besides the model detection rate on the corresponding known attack. As mentioned above, the goal of this scenario is to adapt (\ie, expand, train, and compress) the initial model to new anomalies separately and report the detection rate at different steps (called evaluation states). The second column represents the state of the reported detection rate, and the rest of the columns indicate the detection rate of the model over new (\ie, zero-day) anomalies for three evaluation states\footnote{Note that the goal of this evaluation is to investigate the effectiveness of the updating procedure when the model is faced with new (\ie, zero-day) attacks (\ie, different from its initial known attack). Hence, the experiments where the initial and zero-day attacks are the same are not reported.}: \textit{Before Update (zero-day)},  \textit{After Update (zero-day)}, and  \textit{After Update (initial known)}. 
	
	Prior to adapting the model to a new anomaly, the initial model detection rate is measured on the corresponding zero-day dataset and reported as  \textit{Before Update (zero-day)}. The compressed model detection rate on the same set is reported as  \textit{After Update (zero-day)} to indicate the model's improvement after continual learning. In addition, the  \textit{After Update (initial)} state represents how the updating procedure affects the model's previous knowledge (\ie, catastrophic forgetting) by measuring the compressed model's detection rate on the anomaly which the model was initially trained on.
	
	\begin{table*}[!h]
		\caption{CNN-based model. Detection rate of continual learning for each pair of anomalies, evaluated on the CIC-IDS2017 dataset.}
		\label{CNN:cont-2017}
		\centering
		\resizebox{\textwidth}{!}{
			\begin{tabular}{|c|c|c|c|c|c|c|c|c|c|c|c|c|}
				\hline
				\multicolumn{1}{|l|}{\backslashbox[48mm]{\parbox[c][1.5cm][c]{3cm}{\centering \textbf{Known attack (accuracy)}}}{\parbox[c][1.5cm][c]{1.5cm}{\centering \textbf{Zero-day attack}}  }} & 
				\multicolumn{1}{c|}{\textbf{State}} & \multicolumn{1}{c|}{\textbf{Botnet}} & \multicolumn{1}{c|}{\textbf{DDOS}} & \multicolumn{1}{c|}{\textbf{Portscan}} & \multicolumn{1}{c|}{\textbf{DOS SlowHttpTest}} & \multicolumn{1}{c|}{\textbf{DOS SlowLoris}} & \multicolumn{1}{c|}{\textbf{DOS Hulk}} & \multicolumn{1}{c|}{\textbf{DOS GoldenEye}} & \multicolumn{1}{c|}{\textbf{FTP Patator}} & \multicolumn{1}{c|}{\textbf{SSH Patator}} & \multicolumn{1}{c|}{\textbf{Web BruteForce}} & \multicolumn{1}{c|}{\textbf{Web XSS}} \\ \hline
				& Before Update (zero-day)                             & -           & 0.32                               & 0.33                                   & 0.32                                           & 0.56                                        & 0.32                                   & 0.42                                        & 0.32                                     & 0.32                                     & 0.40                                         & 0.57                                  \\ \cline{2-13} 
				& After Update (zero-day)                              & -           & 0.99                               & 0.99                                   & 0.98                                           & 0.98                                        & 0.99                                   & 0.94                                        & 1.00                                     & 0.97                                     & 0.99                                         & 0.98                                  \\ \cline{2-13} 
				\multirow{-3}{*}{\textbf{Botnet (0.96)}}           & After Update (initial)                            & -           & 0.97                               & 0.97                                   & 0.95                                           & 0.96                                        & 0.96                                   & 0.95                                        & 0.97                                     & 0.95                                     & 0.96                                         & 0.96                                  \\ \hline
				& Before Update (zero-day)                             & 0.45                                 & -         & 0.33                                   & 0.34                                           & 0.33                                        & 0.45                                   & 0.39                                        & 0.33                                     & 0.33                                     & 0.42                                         & 0.59                                  \\ \cline{2-13} 
				& After Update (zero-day)                              & 0.98                                 & -         & 1.00                                   & 0.99                                           & 0.97                                        & 0.90                                   & 0.95                                        & 1.00                                     & 0.99                                     & 0.99                                         & 0.99                                  \\ \cline{2-13} 
				\multirow{-3}{*}{\textbf{DDoS (0.99)}}             & After Update (initial)                           & 0.99                                 & -         & 0.99                                   & 0.99                                           & 0.99                                        & 0.99                                   & 0.98                                        & 0.99                                     & 0.99                                     & 0.98                                         & 0.99                                  \\ \hline
				& Before Update (zero-day)                             & 0.44                                 & 0.33                               & -             & 0.33                                           & 0.58                                        & 0.33                                   & 0.33                                        & 0.33                                     & 0.33                                     & 0.41                                         & 0.58                                  \\ \cline{2-13} 
				& After Update (zero-day)                              & 0.95                                 & 0.98                               & -             & 0.99                                           & 0.93                                        & 0.97                                   & 0.95                                        & 1.00                                     & 0.98                                     & 0.98                                         & 0.98                                  \\ \cline{2-13} 
				\multirow{-3}{*}{\textbf{Portscan (0.99)}}         & After Update (initial)                            & 0.98                                 & 0.99                               & -             & 1.00                                           & 1.00                                        & 1.00                                   & 0.99                                        & 1.00                                     & 0.99                                     & 0.98                                         & 1.00                                  \\ \hline
				& Before Update (zero-day)                             & 0.44                                 & 0.49                               & 0.34                                   & -                     & 0.91                                        & 0.81                                   & 0.63                                        & 0.33                                     & 0.33                                     & 0.43                                         & 0.60                                  \\ \cline{2-13} 
				& After Update (zero-day)                              & 0.97                                 & 0.98                               & 1.00                                   & -                     & 0.99                                        & 0.99                                   & 0.97                                        & 1.00                                     & 0.99                                     & 0.99                                         & 0.98                                  \\ \cline{2-13} 
				\multirow{-3}{*}{\textbf{DoS SlowHttpTest (0.98)}} & After Update (initial)                            & 0.97                                 & 0.99                               & 0.99                                   & -                     & 0.99                                        & 0.97                                   & 0.99                                        & 0.99                                     & 0.99                                     & 0.98                                         & 0.98                                  \\ \hline
				& Before Update (zero-day)                             & 0.44                                 & 0.33                               & 1.00                                   & 0.35                                           & -                  & 0.33                                   & 0.35                                        & 0.33                                     & 0.33                                     & 0.41                                         & 0.58                                  \\ \cline{2-13} 
				& After Update (zero-day)                              & 0.97                                 & 0.98                               & 1.00                                   & 0.96                                           & -                  & 0.99                                   & 0.92                                        & 1.00                                     & 0.98                                     & 0.99                                         & 0.99                                  \\ \cline{2-13} 
				\multirow{-3}{*}{\textbf{DoS SlowLoris (0.99)}}    & After Update (initial)                            & 0.98                                 & 0.98                               & 0.99                                   & 0.98                                           & -                  & 0.97                                   & 0.93                                        & 0.98                                     & 0.96                                     & 0.97                                         & 0.96                                  \\ \hline
				& Before Update (zero-day)                             & 0.44                                 & 0.72                               & 0.33                                   & 0.45                                           & 0.34                                        & -             & 0.84                                        & 0.33                                     & 0.33                                     & 0.42                                         & 0.59                                  \\ \cline{2-13} 
				& After Update (zero-day)                              & 0.96                                 & 0.99                               & 1.00                                   & 0.98                                           & 0.98                                        & -             & 0.97                                        & 1.00                                     & 0.99                                     & 0.99                                         & 0.99                                  \\ \cline{2-13} 
				\multirow{-3}{*}{\textbf{DoS Hulk (0.97)}}       & After Update (initial)                            & 0.98                                 & 0.99                               & 0.98                                   & 0.99                                           & 0.99                                        & -             & 0.98                                        & 0.99                                     & 0.99                                     & 0.98                                         & 0.99                                  \\ \hline
				& Before Update (zero-day)                             & 0.72                                 & 0.77                               & 0.33                                   & 0.42                                           & 0.65                                        & 0.99                                   & -                  & 0.33                                     & 0.33                                     & 0.44                                         & 0.59                                  \\ \cline{2-13} 
				& After Update (zero-day)                              & 0.97                                 & 0.98                               & 1.00                                   & 0.98                                           & 0.97                                        & 0.99                                   & -                  & 1.00                                     & 0.99                                     & 1.00                                         & 0.99                                  \\ \cline{2-13} 
				\multirow{-3}{*}{\textbf{DoS GoldenEye (0.98)}}    & After Update (initial)                            & 0.99                                 & 0.98                               & 0.97                                   & 0.99                                           & 0.97                                        & 0.98                                   & -                  & 0.99                                     & 0.99                                     & 0.98                                         & 0.99                                  \\ \hline
				& Before Update (zero-day)                             & 0.45                                 & 0.33                               & 0.33                                   & 0.33                                           & 0.33                                        & 0.33                                   & 0.33                                        & -               & 0.33                                     & 0.42                                         & 0.59                                  \\ \cline{2-13} 
				& After Update (zero-day)                              & 0.76                                 & 0.82                               & 1.00                                   & 0.97                                           & 0.97                                        & 0.91                                   & 0.95                                        & -               & 0.99                                     & 0.98                                         & 0.98                                  \\ \cline{2-13} 
				\multirow{-3}{*}{\textbf{FTP Patator (0.99)}}       & After Update (initial)                            & 1.00                                 & 1.00                               & 1.00                                   & 1.00                                           & 0.98                                        & 1.00                                   & 0.98                                        & -               & 1.00                                     & 0.99                                         & 0.99                                  \\ \hline
				& Before Update (zero-day)                             & 0.45                                 & 0.33                               & 0.33                                   & 0.33                                           & 0.33                                        & 0.33                                   & 0.33                                        & 0.33                                     & -               & 0.42                                         & 0.59                                  \\ \cline{2-13} 
				& After Update (zero-day)                              & 0.97                                 & 0.99                               & 0.99                                   & 0.98                                           & 0.97                                        & 0.98                                   & 0.95                                        & 1.00                                     & -               & 1.00                                         & 0.98                                  \\ \cline{2-13} 
				\multirow{-3}{*}{\textbf{SSH Patator (0.99)}}       & After Update (initial)                            & 0.99                                 & 0.99                               & 0.99                                   & 1.00                                           & 0.99                                        & 0.98                                   & 0.99                                        & 1.00                                     & -               & 0.99                                         & 0.99                                  \\ \hline
				& Before Update (zero-day)                             & 0.44                                 & 0.33                               & 0.33                                   & 0.33                                           & 0.33                                        & 0.33                                   & 0.33                                        & 0.33                                     & 0.33                                     & -                   & 0.98                                  \\ \cline{2-13} 
				& After Update (zero-day)                              & 0.96                                 & 0.98                               & 1.00                                   & 0.97                                           & 0.93                                        & 0.97                                   & 0.94                                        & 1.00                                     & 0.99                                     & -                   & 0.99                                  \\ \cline{2-13} 
				\multirow{-3}{*}{\textbf{BruteForce Web (0.97)}}   & After Update (initial)                            & 0.95                                 & 0.99                               & 0.99                                   & 0.99                                           & 0.99                                        & 0.99                                   & 0.98                                        & 0.99                                     & 0.99                                     & -                   & 0.99                                  \\ \hline
				& Before Update (zero-day)                             & 0.66                                 & 0.32                               & 0.32                                   & 0.76                                           & 0.33                                        & 0.32                                   & 0.32                                        & 0.32                                     & 0.32                                     & 0.92                                         & -            \\ \cline{2-13} 
				& After Update (zero-day)                              & 0.91                                 & 0.98                               & 0.98                                   & 0.99                                           & 0.90                                        & 0.98                                   & 0.97                                        & 0.99                                     & 0.99                                     & 0.98                                         & -            \\ \cline{2-13} 
				\multirow{-3}{*}{\textbf{XSS Web (0.94)}}          & After Update (initial)                            & 0.98                                 & 0.99                               & 0.97                                   & 0.99                                           & 0.96                                        & 0.98                                   & 0.98                                        & 0.98                                     & 0.99                                     & 0.97                                         & -            \\ \hline
		\end{tabular}}
	\end{table*}
	
	\npdecimalsign{.}
	\nprounddigits{2}

	\begin{table*}[!h]
		\caption{LSTM-based model. Detection rate of continual learning for each pair of anomalies, evaluated on the CIC-IDS2017 dataset.}
		\label{LSTM:cont-2017}
		\centering
		\resizebox{\textwidth}{!}{
			
			\begin{tabular}{|c|c|c|c|c|c|c|c|c|c|c|c|c|}
				\hline
				\multicolumn{1}{|l|} {\backslashbox[48mm]{\parbox[c][1.5cm][c]{3cm}{\centering \textbf{Known attack (accuracy)}}}{\parbox[c][1.5cm][c]{1.5cm}{\centering \textbf{Zero-day attack}}  }}                      & \multicolumn{1}{c|}{\textbf{State}} & \multicolumn{1}{c|}{\textbf{Botnet}} & \multicolumn{1}{c|}{\textbf{DDoS}} & \multicolumn{1}{c|}{\textbf{Portscan}} & \multicolumn{1}{c|}{\textbf{DoS SlowHttpTest}} & \multicolumn{1}{c|}{\textbf{DoS SlowLoris}} & \multicolumn{1}{c|}{\textbf{DoS Hulk}} & \multicolumn{1}{c|}{\textbf{DoS GoldenEye}} & \multicolumn{1}{c|}{\textbf{FTP Patator}} & \multicolumn{1}{c|}{\textbf{SSH Patator}} & \multicolumn{1}{c|}{\textbf{BruteForce Web}} & \multicolumn{1}{c|}{\textbf{XSS Web}} \\ \hline
				& Before Update (zero-day)                     & -           & 0.31                           & 0.31                               & 0.31                                       & 0.31                                    & 0.32                               & 0.45                                    & 0.31                                 & 0.31                                 & 0.43                                     & 0.57                              \\ \cline{2-13} 
				& After Update (zero-day)                      & -           & 0.93                              & 0.95                               & 0.94                                       & 0.92                                    & 0.92                               & 0.93                                    & 0.93                                 & 0.95                                 & 0.91                                     & 0.87                              \\ \cline{2-13} 
				\multirow{-3}{*}{\textbf{Botnet (0.93)}}           & After Update (initial)                        & -           & 0.91                            & 0.90                               & 0.92                                       & 0.90                                    & 0.90                               & 0.91                                    & 0.91                                 & 0.94                                 & 0.91                                     & 0.90                              \\ \hline
				& Before Update (zero-day)                     & 0.69                             & -         & 0.29                                   & 0.44                                       & 0.30                                       & 0.77                               & 0.83                                    & 0.29                                     & 0.95                                 & 0.40                                     & 0.52                              \\ \cline{2-13} 
				& After Update (zero-day)                      & 0.90                             & -         & 0.95                               & 0.78                                       & 0.81                                    & 0.91                               & 0.89                                    & 0.90                                 & 0.96                                 & 0.87                                     & 0.87                                 \\ \cline{2-13} 
				\multirow{-3}{*}{\textbf{DDoS (0.92)} }             & After Update (initial)                        & 0.93                                & -         & 0.91                               & 0.92                                       & 0.89                                    & 0.93                               & 0.93                                    & 0.88                                    & 0.93                                 & 0.90                                     & 0.90                              \\ \hline
				& Before Update (zero-day)                     & 0.63                             & 0.50                             & -             & 0.50                                        & 0.50                                         & 0.49                                 & 0.50                                      & 0.50                                   & 0.49                                   & 0.59                                     & 0.74                              \\ \cline{2-13} 
				& After Update (zero-day)                      & 0.64                            & 0.50                             & -             & 0.62                                           & 0.55                                      & 0.50                                 & 0.50                                      & 0.50                                   & 0.50                                   & 0.90                                     & 0.93                             \\ \cline{2-13} 
				\multirow{-3}{*}{\textbf{Portscan (0.98)}}         & After Update (initial)                        & 1.00                              & 1.00                            & -             & 0.97                                           & 1.00                                     & 1.00                                & 1.00                                     & 1.00                                  & 1.00                                  & 0.96                                        & 0.97                               \\ \hline
				& Before Update (zero-day)                     & 0.51                             & 0.57                           & 0.34                                  & -                     & 0.75                                    & 0.43                                  & 0.40                                    & 0.78                                 & 0.33                                 & 0.79                                     & 0.82                              \\ \cline{2-13} 
				& After Update (zero-day)                      & 0.80                             & 0.94                           & 0.98                               & -                     & 0.75                                    & 0.43                                  & 0.77                                    & 0.99                                 & 0.92                                    & 0.92                                     & 0.96                              \\ \cline{2-13} 
				\multirow{-3}{*}{\textbf{DoS SlowHttpTest (0.98)} } & After Update (initial)                        & 0.86                             & 0.97                               & 0.98                               & -                     & 0.99                                        & 0.99                                   & 0.87                                    & 0.99                                    & 0.98                                 & 0.99                                        & 0.99                                 \\ \hline
				& Before Update (zero-day)                     & 0.45                             & 0.33                           & 0.38                               & 0.84                                       & -                  & 0.44                               & 0.40                                    & 0.34                                    & 0.33                                 & 0.42                                     & 0.59                              \\ \cline{2-13} 
				& After Update (zero-day)                      & 0.73                              & 0.66                           & 0.82                               & 0.93                                       & -                  & 0.94                                   & 0.84                                        & 0.77                                    & 0.63                                 & 0.77                                     & 0.60                              \\ \cline{2-13} 
				\multirow{-3}{*}{\textbf{DoS SlowLoris (0.97)}}    & After Update (initial)                        & 0.95                             & 0.93                           & 0.97                                   & 0.98                                       & -                  & 0.97                                   & 0.98                                    & 0.97                                    & 0.93                                 & 0.96                                     & 0.99                                  \\ \hline
				& Before Update (zero-day)                     & 0.45                             & 0.34                              & 0.33                               & 0.36                                       & 0.43                                    & -             & 0.60                                    & 0.36                                     & 0.33                                 & 0.43                                     & 0.59                              \\ \cline{2-13} 
				& After Update (zero-day)                      & 0.92                             & 0.92                           & 0.97                               & 0.97                                           & 0.97                                    & -             & 0.95                                    & 0.99                                 & 0.98                                 & 0.92                                     & 0.92                              \\ \cline{2-13} 
				\multirow{-3}{*}{\textbf{DoS Hulk (0.99)}}         & After Update (initial)                        & 0.97                                & 0.96                           & 0.96                               & 0.97                                           & 0.99                                    & -             & 0.98                                    & 0.99                                 & 0.97                                 & 0.96                                     & 0.97                              \\ \hline
				& Before Update (zero-day)                     & 0.48                             & 0.97                           & 0.33                               & 0.40                                       & 0.67                                    & 0.99                                   & -                  & 0.33                                 & 0.33                                 & 0.42                                     & 0.60                              \\ \cline{2-13} 
				& After Update (zero-day)                      & 0.74                             & 1.00                           & 0.98                                   & 0.95                                           & 0.91                                        & 0.99                               & -                  & 1.00                                 & 0.93                                 & 0.94                                     & 0.96                               \\ \cline{2-13} 
				\multirow{-3}{*}{\textbf{DoS GoldenEye (0.99)}}    & After Update (initial)                        & 0.99                             & 0.99                           & 0.98                                  & 0.99                                           & 0.98                                       & 0.99                               & -                  & 0.99                                 & 0.93                                 & 0.95                                     & 0.98                              \\ \hline
				& Before Update (zero-day)                     & 0.45                             & 0.33                           & 0.45                                   & 0.33                                       & 0.34                                    & 0.33                               & 0.33                                    & -               & 0.33                                 & 0.42                                     & 0.60                              \\ \cline{2-13} 
				& After Update (zero-day)                      & 0.81                             & 0.39                              & 0.99                               & 0.75                                          & 0.54                                    & 0.33                                   & 0.33                                    & -               & 0.98                                 & 0.91                                     & 0.93                              \\ \cline{2-13} 
				\multirow{-3}{*}{\textbf{FTP Patator (0.99)}}       & After Update (initial)                        & 0.96                             & 0.98                           & 1.00                                  & 0.99                                       & 1.00                                    & 1.00                               & 1.00                                           & -               & 1.00                                    & 0.98                                     & 0.98                              \\ \hline
				& Before Update (zero-day)                     & 0.49                              & 0.80                           & 0.87                               & 0.55                                          & 0.37                                    & 0.29                               & 0.29                                    & 0.92                                 & -               & 0.81                                     & 0.83                              \\ \cline{2-13} 
				& After Update (zero-day)                      & 0.59                             & 0.92                              & 0.92                                  & 0.77                                          & 0.68                                        & 0.37                                  & 0.33                                    & 0.92                                 & -               & 0.81                                     & 0.83                              \\ \cline{2-13} 
				\multirow{-3}{*}{\textbf{SSH Patator (0.92))}}       & After Update (initial)                        & 0.91                                & 0.91                              & 0.91                                  & 0.91                                          & 0.91                                       & 0.91                                  & 0.91                                       & 0.92                                 & -               & 0.92                                     & 0.92                              \\ \hline
				& Before Update (zero-day)                     & 0.48                             & 0.32                           & 0.36                                  & 0.41                                       & 0.57                                    & 0.31                               & 0.31                                    & 0.33                                    & 0.31                                 & -                   & 0.96                               \\ \cline{2-13} 
				& After Update (zero-day)                      & 0.92                             & 0.81                           & 0.98                               & 0.85                                       & 0.65                                        & 0.72                               & 0.71                                    & 0.97                                 & 0.81                                     & -                   & 0.95                              \\ \cline{2-13} 
				\multirow{-3}{*}{\textbf{BruteForce Web (0.95)}}   & After Update (initial)                        & 0.95                             & 0.89                           & 0.97                               & 0.93                                       & 0.93                                    & 0.90                              & 0.93                                    & 0.96                                  & 0.94                                 & -                   & 0.97                              \\ \hline
				& Before Update (zero-day)                     & 0.65                             & 0.62                           & 0.98                               & 0.88                                       & 0.63                                    & 0.33                               & 0.36                                        & 0.98                                 & 0.31                                     & 0.94                                     & -            \\ \cline{2-13} 
				& After Update (zero-day)                      & 0.89                              & 0.92                               & 0.99                                  & 0.91                                       & 0.91                                    & 0.89                                  & 0.85                                    & 0.99                                    & 0.98                                 & 0.95                                     & -            \\ \cline{2-13} 
				\multirow{-3}{*}{\textbf{XSS Web (0.95)}}          & After Update (initial)                        & 0.94                               & 0.94                           & 0.97                               & 0.97                                       & 0.87                                    & 0.89                               & 0.91                                    & 0.97                                 & 0.96                                 & 0.97                                     & -              \\ \hline
		\end{tabular}}
	\end{table*}

	In the second scenario, similar to the first one, an initial model is trained on a known anomaly. Then, considering the rest of the anomalies as zero-day attacks, the model is sequentially expanded, trained, and compressed on 128 flows of each of the remaining new attacks. The main difference between the first and second scenarios is that the latter uses the previous step's compressed model as the initial model for the current training step. In other words, during the continual learning procedure, the model acquires knowledge about all the previous anomalies and considers them as known attacks. 
	
	Similar to the previous scenario, we perform different evaluation experiments. In each experiment, we use a different permutation for the attack sequence. Finally, the detection rate of each step is reported according to the average detection rate of all experiments' corresponding steps. As a result, this scenario does not rely on a particular attack sequence and yields more reliable results for real-world situations. 
	
	In order to evaluate the second scenario in the test phase, we prepare two datasets for each experiment's step. The first one, called \emph{zero-day dataset}, includes 500 new attack flows and 500 benign flows. The second one, named as the \emph{known dataset}, consists of 500 attack flows for each previously known attack in addition to an equal number of benign flows for making the dataset balanced. Notice that the known dataset expands as the evaluation steps progress over the attack sequence.

\begin{figure}[!ht]
		\centering
		\includegraphics[width= 0.5\linewidth]{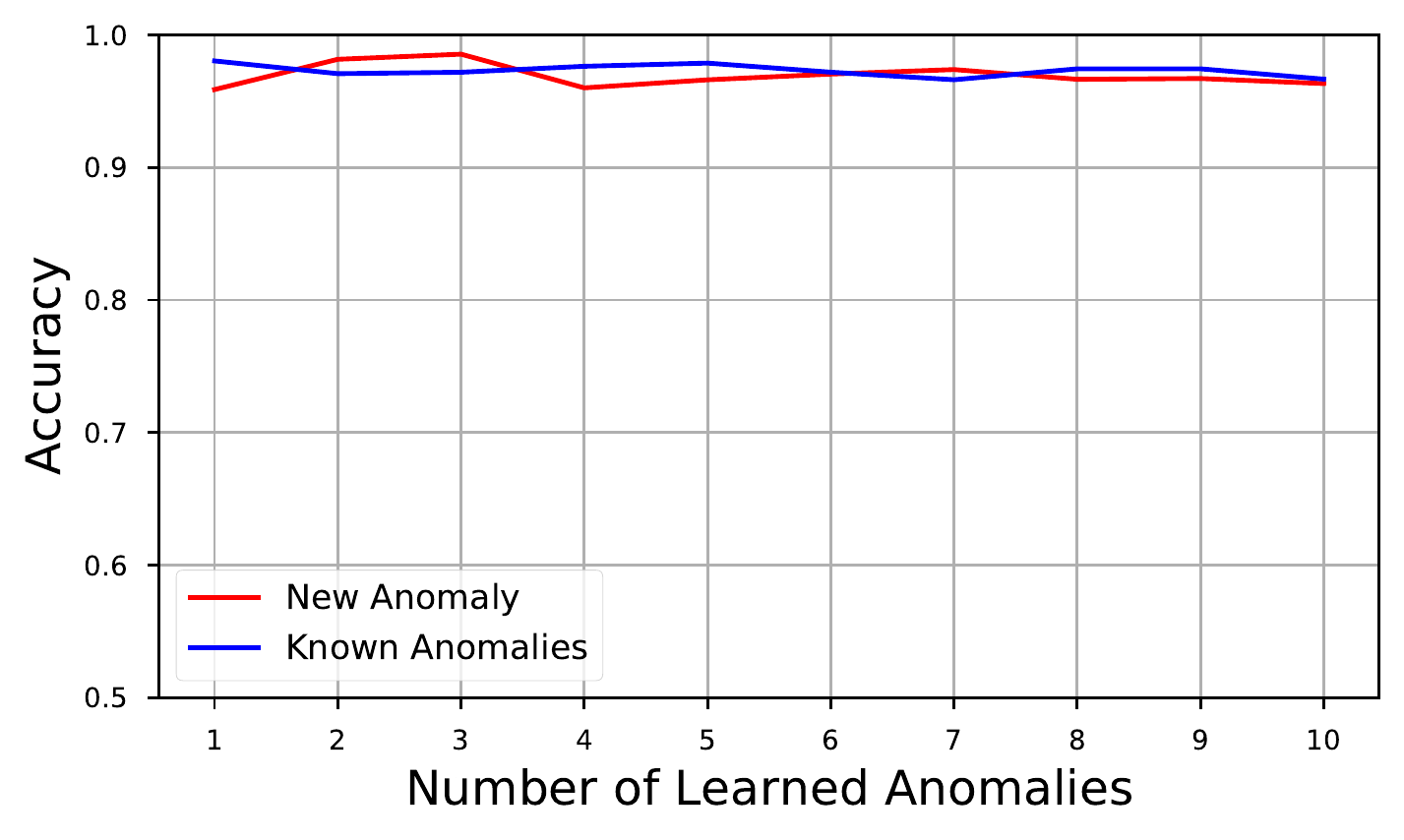}
		\caption{CNN-Based model detection rate after each step of learning a new anomaly on the CIC-IDS2017 dataset.}
		\label{fig:cnncont2017}
	\end{figure}

\begin{figure}[!ht]
		\centering
		\includegraphics[width=0.5\linewidth]{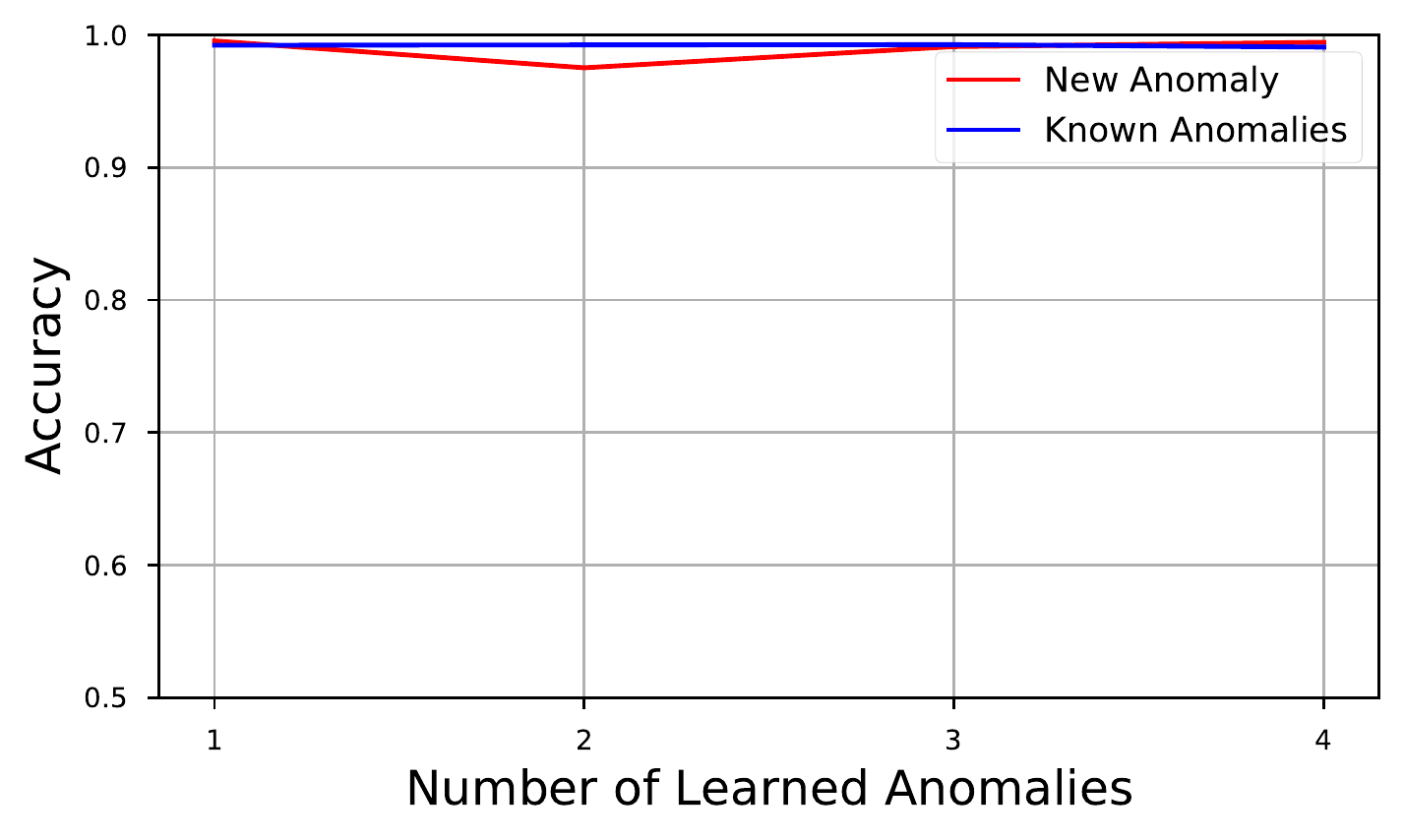}
		\caption{CNN-Based model detection rate after each step of learning a new anomaly on the CSE-CIC-IDS2018 dataset.}
		\label{fig:cnncont2018}
	\end{figure}

\begin{figure}[!ht]
		\centering
		\includegraphics[width=0.5\linewidth]{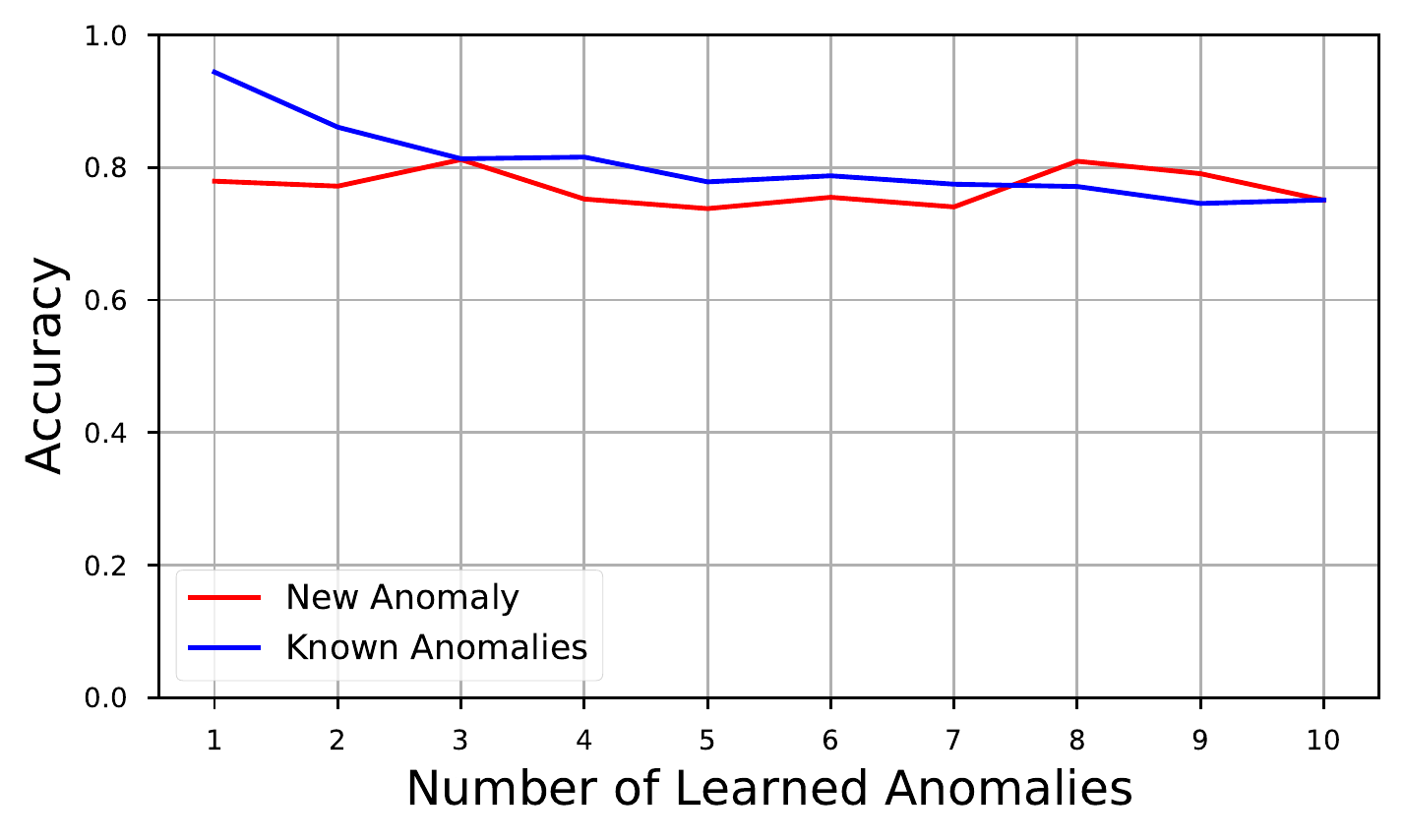}
		\caption{LSTM-Based model detection rate after each step of learning a new anomaly on the CIC-IDS2017 dataset.}
		\label{fig:lstmcont2017}
	\end{figure}

\begin{figure}[!ht]
		\centering
		\includegraphics[width=0.5\linewidth]{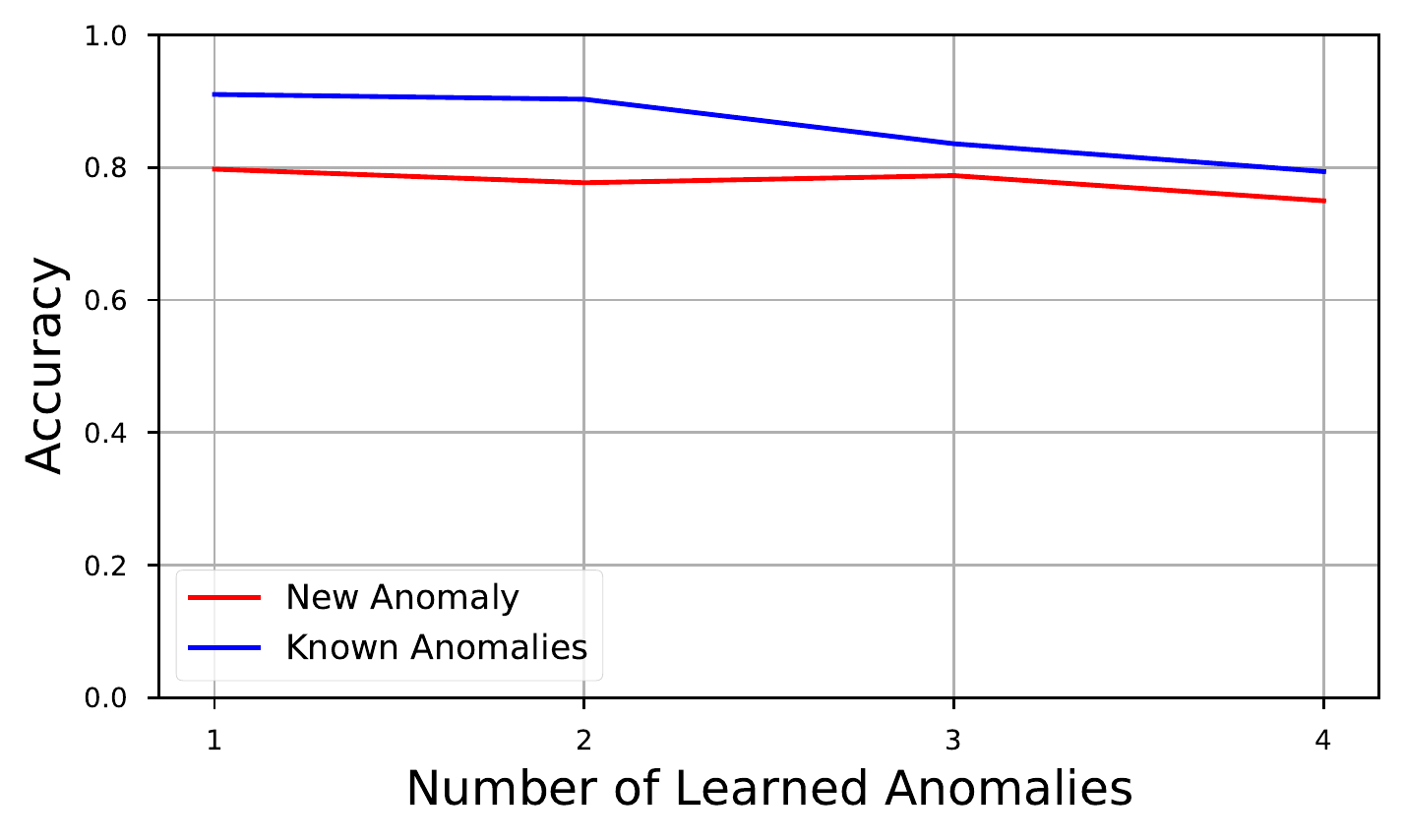}
		\caption{LSTM-Based model detection rate after each step of learning a new anomaly on the CSE-CIC-IDS2018 dataset.}
		\label{fig:lstmcont2018}
	\end{figure}

	Figures \ref{fig:cnncont2017} and \ref{fig:cnncont2018} depict the results of this experiment with CNN-based models over the CIC-IDS2017 and CSE-CIC-IDS2018 datasets, respectively. Similarly, Figures \ref{fig:lstmcont2017} and \ref{fig:lstmcont2018} report the results on the same datasets with LSTM-based models.
	The results indicate that while the proposed adaptive deep IDS can continually adapt itself to the new zero-day attacks, it also preserves its ability to detect the previously observed attacks. Furthermore, the CNN-based models have a better average detection rate than LSTM-based models for detecting new anomalies (we will discuss more about the reasons of the different results produced by CNN and LSTM models in Section \ref{sec:discussion}). To be more precise, the CNN-based models have an average detection rate above 95\% both on new and previously known attacks (\ie, after the updating procedure). On the other hand, LSTM-based models tend to have a lower detection rate when updated on new attacks. However, it is worth mentioning that their previous knowledge is preserved during the updating procedure (\ie, the detection rate on known anomalies does not decrease after learning a new attack).

	\subsection{Federated Learning}
	\label{sec:expfed}
	As discussed in Section~\ref{sec:framework_federated}, the federated learning technique is essential to a distributed DL-based IDS. In this section, we aim to evaluate the performance of the federated learning implementation of our proposed framework.
	
	Although agents often may have encountered benign or known attacks in practice, we consider a more challenging case in which each agent analyzes a completely new zero-day attack for evaluating the proposed multi-agent architecture.
	In this scenario, the main-model is initially trained on an anomaly as the known attack. Then, a process thread is designated as an agent for each of the remaining anomalies. Each agent is responsible for learning a new anomaly and updating the main-model. When this (simultaneous) learning and (asynchronous) updating process is done, the performance of the final version of the main-model is evaluated and reported in Tables \ref{tab:CNN-fed-2017} to \ref{tab:LSTM-fed-2018}. Also, in our experiments, we set $\alpha$ in \eqref{eq:fisher-upd} as the ratio between the number of samples used in training the main-model and each sub-model.
	
	The evaluation procedure is similar to Section \ref{sec:DeepAdaptive}. The main difference is that the zero-day dataset comprises a collective set consisting of 500 flows from each zero-day anomaly and a proportionate amount of benign flows. Consequently, the \emph{Unknowns-After} state represents the model detection rate on all the unknown attacks after the federated updating phase.
	
	\begin{table}[!ht]
		\caption{CNN-based model detection rate in the federated learning approach on the CIC-IDS2017 dataset.} 
		\label{tab:CNN-fed-2017}
		\centering
		\resizebox{0.6\linewidth}{!}{
			\begin{tabular}{|c|c|c|c|}
				\hline
				\multicolumn{1}{|l|}{\backslashbox{\parbox[c][1cm][c]{1cm}{\centering \textbf{Known attack}}}{\parbox[c][1cm][c]{1cm}{\centering \textbf{State}}  }}      & \multicolumn{1}{c|}{\textbf{Unknowns-Before}} & \multicolumn{1}{c|}{\textbf{Unknowns-After}} & \multicolumn{1}{c|}{\textbf{Known-After}} \\ \hline
				\textbf{Botnet}             & 0.49                                       & 0.96                                                & 0.95                                                       \\ \hline
				\textbf{DDoS}               & 0.49                                           & 0.96                                                & 0.95                                                      \\ \hline
				\textbf{Portscan}           & 0.51                                       & 0.92                                                & 0.92                                                       \\ \hline
				\textbf{DoS   SlowHttpTest} & 0.65                                       & 0.97                                                & 0.95                                                       \\ \hline
				\textbf{DoS SlowLoris}      & 0.54                                      & 0.96                                                & 0.96                                                      \\ \hline
				\textbf{DoS Hulk}           & 0.58                                       & 0.95                                                & 0.95                                                      \\ \hline
				\textbf{DoS GoldenEye}      & 0.63                                       & 0.96                                                & 0.95                                                      \\ \hline
				\textbf{FTP Patator}         & 0.49                                       & 0.92                                                & 0.91                                                      \\ \hline
				\textbf{SSH Patator}         & 0.49                                       & 0.95                                                & 0.95                                                      \\ \hline
				\textbf{Web BruteForce}     & 0.49                                       & 0.94                                                & 0.94                                                    \\ \hline
				\textbf{Web XSS}            & 0.61                                       & 0.90                                                & 0.91                                                    \\ \hline
		\end{tabular}}
	\end{table}
	
	\begin{table}[!ht]
		\caption{CNN-based model detection rate in the federated learning approach on the CSE-CIC-IDS2018 dataset.}
		\label{tab:CNN-fed-2018}
		\centering
		\resizebox{0.6\linewidth}{!}{
			\begin{tabular}{|c|c|c|c|}
				\hline
				\multicolumn{1}{|l|}{\backslashbox{\parbox[c][1cm][c]{1cm}{\centering \textbf{Known attack}}}{\parbox[c][1cm][c]{1cm}{\centering \textbf{State}}  }}     & \multicolumn{1}{c|}{\textbf{Unknowns-Before}} & \multicolumn{1}{c|}{\textbf{Unknowns-After}} & \multicolumn{1}{c|}{\textbf{Known-After}} \\ \hline
				\textbf{Botnet}            & 0.49                                         & 0.97                                                  & 0.99                                                    \\ \hline
				\textbf{DoS SlowLoris}     & 0.48                                          & 0.99                                                  & 1.00                                                       \\ \hline
				\textbf{DoS GoldenEye}     & 0.58                                         & 0.99                                                 & 0.99                                                    \\ \hline
				\textbf{FTP BruteForce}    & 0.49                                         & 0.98                                                  & 1.00                                                    \\ \hline
				\textbf{SSH   BruteForce} & 0.49                                          & 0.98                                                   & 1.00                                                    \\ \hline
		\end{tabular}}
	\end{table}
	
	\begin{table}[!ht]
		\caption{LSTM-based model detection rate in the federated learning approach on the CIC-IDS2017 dataset.}
		\label{tab:LSTM-fed-2017}
		\centering
		\resizebox{0.6\linewidth}{!}{
			\begin{tabular}{|c|c|c|c|}
				\hline
				\multicolumn{1}{|l|}{\backslashbox{\parbox[c][1cm][c]{1cm}{\centering \textbf{Known attack}}}{\parbox[c][1cm][c]{1cm}{\centering \textbf{State}}  }}      & \multicolumn{1}{c|}{\textbf{Unknowns-Before}} & \multicolumn{1}{c|}{\textbf{Unknowns-After}} & \multicolumn{1}{c|}{\textbf{Knowns-After}} \\ \hline
				\textbf{Botnet}             & 0.69                                      & 0.78                                      & 0.87                                   \\ \hline
				\textbf{DDoS}               & 0.64                                      & 0.83                                     & 0.90                                   \\ \hline
				\textbf{Portscan}           & 0.48                                          & 0.50                                          & 0.98                                       \\ \hline
				\textbf{DoS   SlowHttpTest} & 0.61                                      & 0.61                                     & 0.87                                   \\ \hline
				\textbf{DoS SlowLoris}      & 0.54                                         & 0.90                                     & 0.90                                   \\ \hline
				\textbf{DoS Hulk}           & 0.79                                          & 0.89                                     & 0.89                                   \\ \hline
				\textbf{DoS GoldenEye}      & 0.64                                      & 0.90                                     & 0.92                                   \\ \hline
				\textbf{FTP Patator}         & 0.74                                      & 0.74                                     & 0.89                                   \\ \hline
				\textbf{SSH Patator}         & 0.40                                      & 0.62                                     & 0.78                                   \\ \hline
				\textbf{BruteForce Web}     & 0.80                                      & 0.81                                     & 0.94                                     \\ \hline
				\textbf{XSS Web}            & 0.64                                      & 0.72                                     & 0.91                                      \\ \hline
		\end{tabular}}
	\end{table}
	
	\begin{table}[!ht]
		\caption{LSTM-based model detection rate in the federated learning approach on the CSE-CIC-IDS2018 dataset.}
		\label{tab:LSTM-fed-2018}
		\centering
		\resizebox{0.6\linewidth}{!}{
			\begin{tabular}{|c|c|c|c|}
				\hline
				\multicolumn{1}{|l|}{\backslashbox{\parbox[c][1cm][c]{1cm}{\centering \textbf{Known attack}}}{\parbox[c][1cm][c]{1cm}{\centering \textbf{State}}  }}     & \multicolumn{1}{c|}{\textbf{Unknowns-Before}} & \multicolumn{1}{c|}{\textbf{Unknowns-After}} & \multicolumn{1}{c|}{\textbf{Knowns-After}} \\ \hline
				\textbf{Botnet}            &\hfil0.51                                      & 0.59                                       & 0.98                                   \\ \hline
				\textbf{DoS SlowLoris}     & 0.70                                       & 0.80                                     & 0.98                                   \\ \hline
				\textbf{DoS GoldenEye}     & 0.64                                          & 0.80                                       & 0.98                                       \\ \hline
				\textbf{FTP BruteForce}    & 0.52                                        & 0.62                                       & 0.99                                   \\ \hline
				\textbf{SSH   BruteForce} & 0.48                                          & 0.5                                          & 0.91                                       \\ \hline
		\end{tabular}}
	\end{table}

	\subsection{Early Attack Detection Through Packet Assessment}
	\label{sec:exppacket}
	This section evaluates an LSTM model's ability to gradually assign a probability to each packet of an incoming flow. We consider a many-to-many LSTM-based model with the same architecture described in Section \ref{sec:architecture} and train it on a collection of all the anomalies in the CIC-IDS2017 dataset. The model yields an anomaly probability per input packet. Finally, we have an output vector whose size equals the number of packets in the incoming flow. 
	
\begin{figure*}[!h]
		\centering
		\subfloat[Botnet]{\includegraphics[width=.3\textwidth]{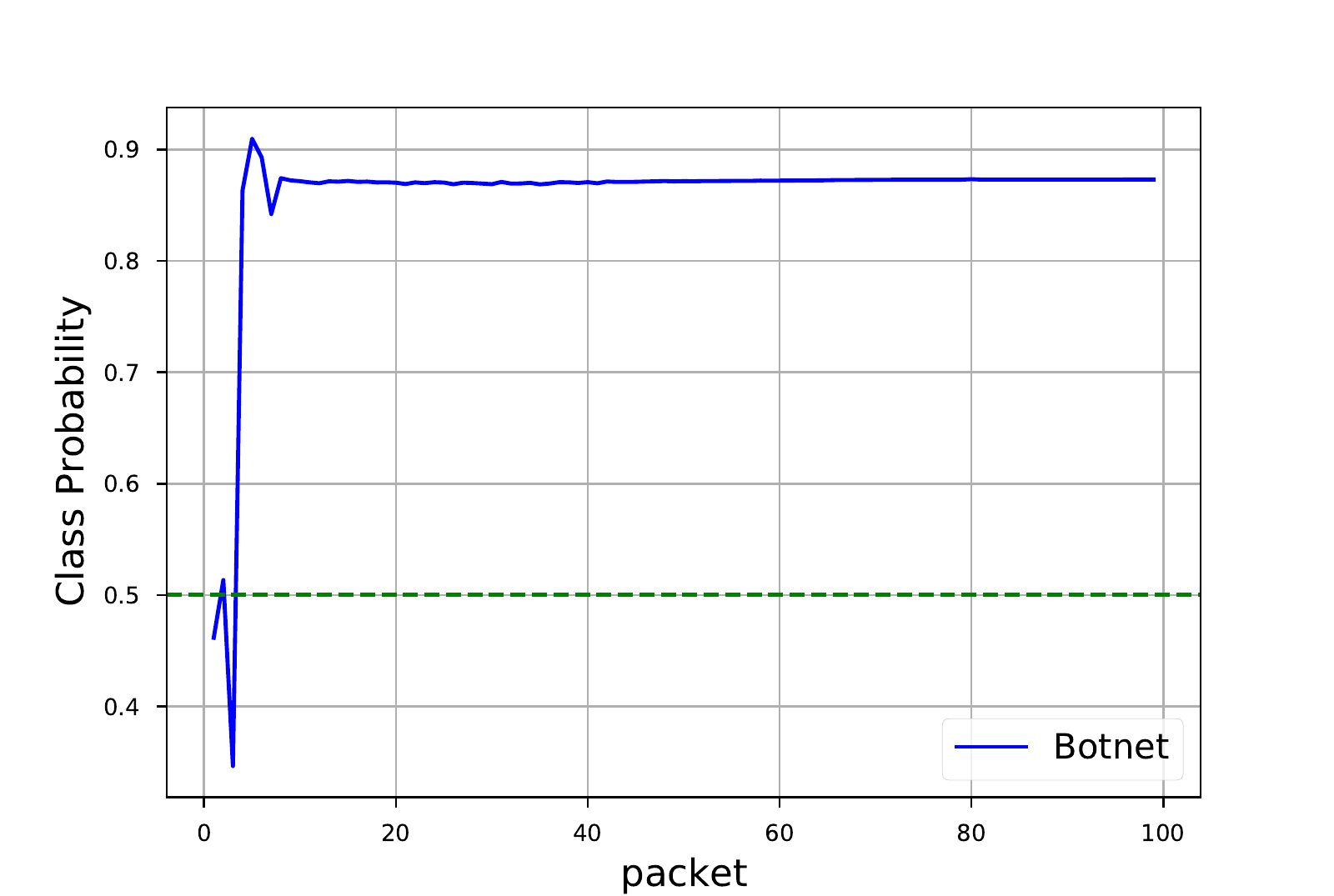}}
		\subfloat[DDoS]{\includegraphics[width=.3\textwidth]{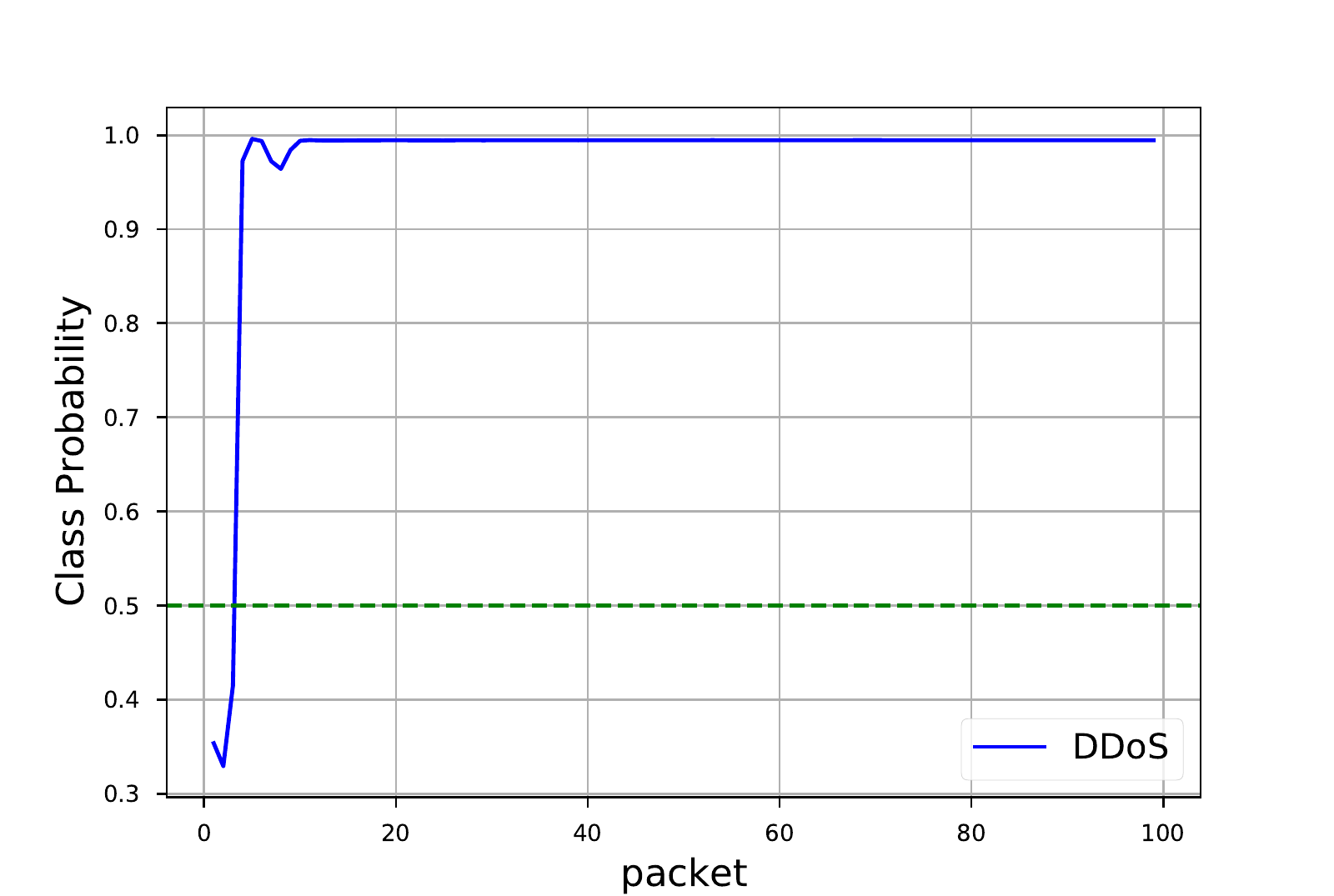}}
		\subfloat[Portscan]{\includegraphics[width=.3\textwidth]{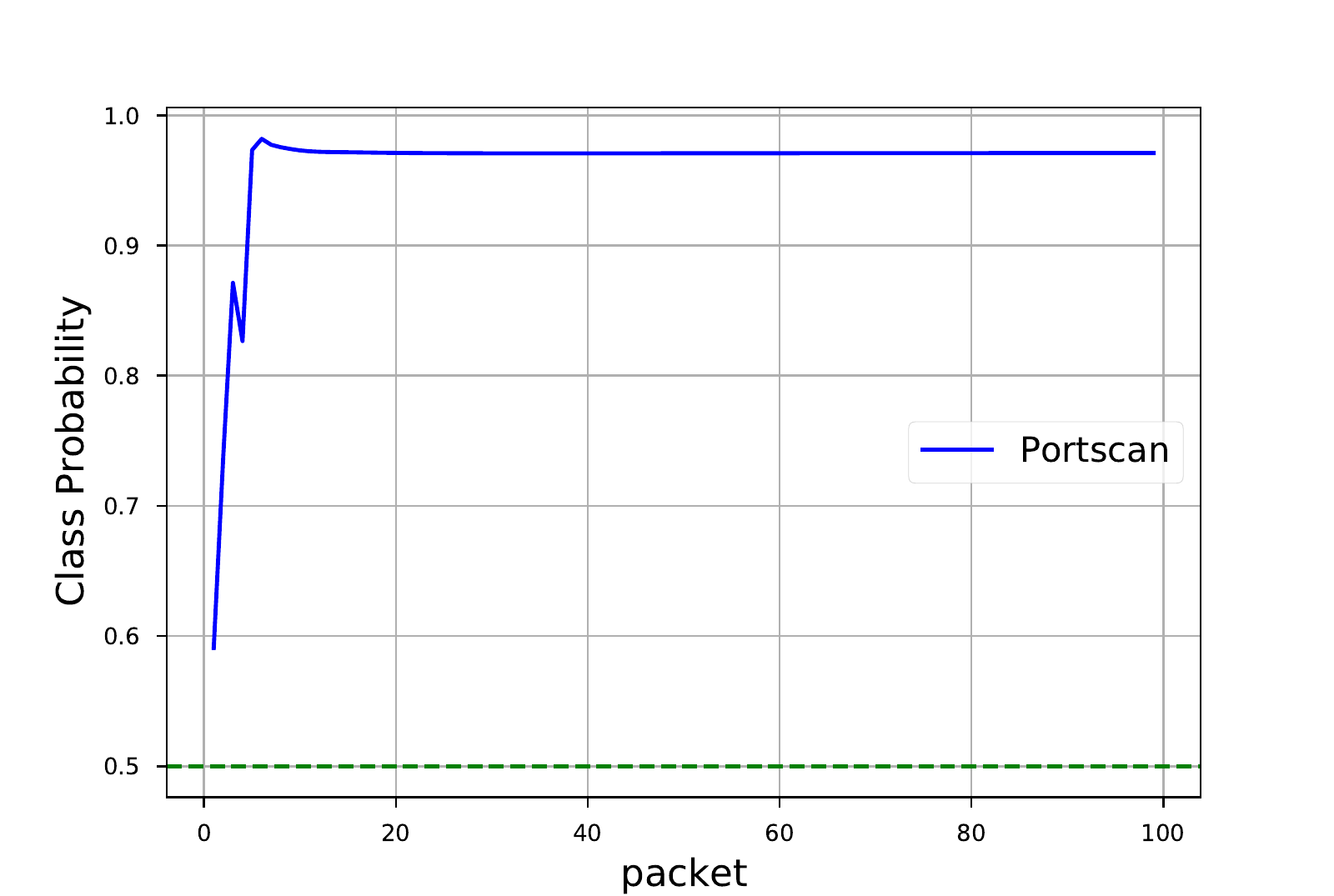}}
		\hfill
		\subfloat[DoS SlowHttpTest]{\includegraphics[width=.3\textwidth]{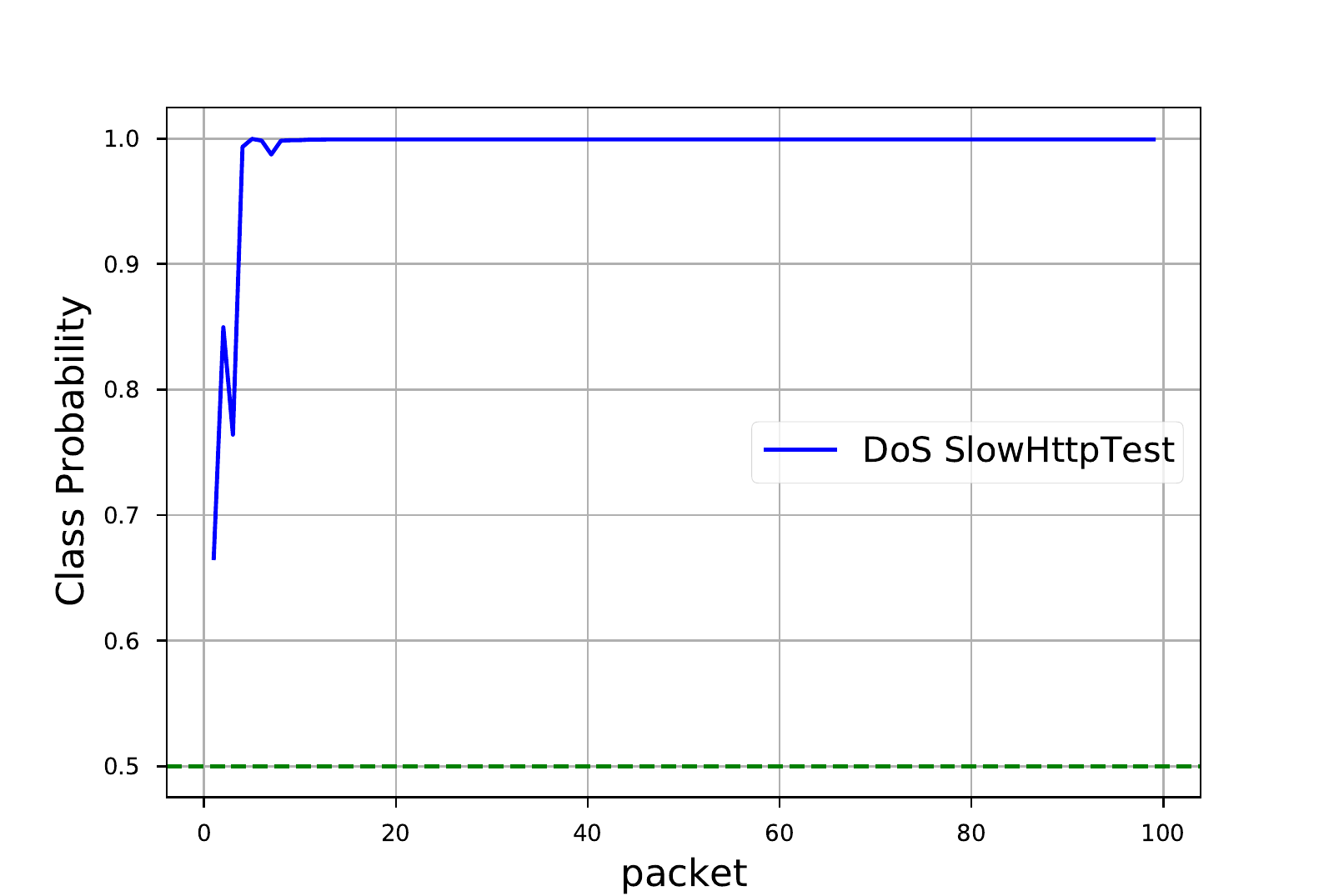}}
		\subfloat[DoS SlowLoris]{\includegraphics[width=.3\textwidth]{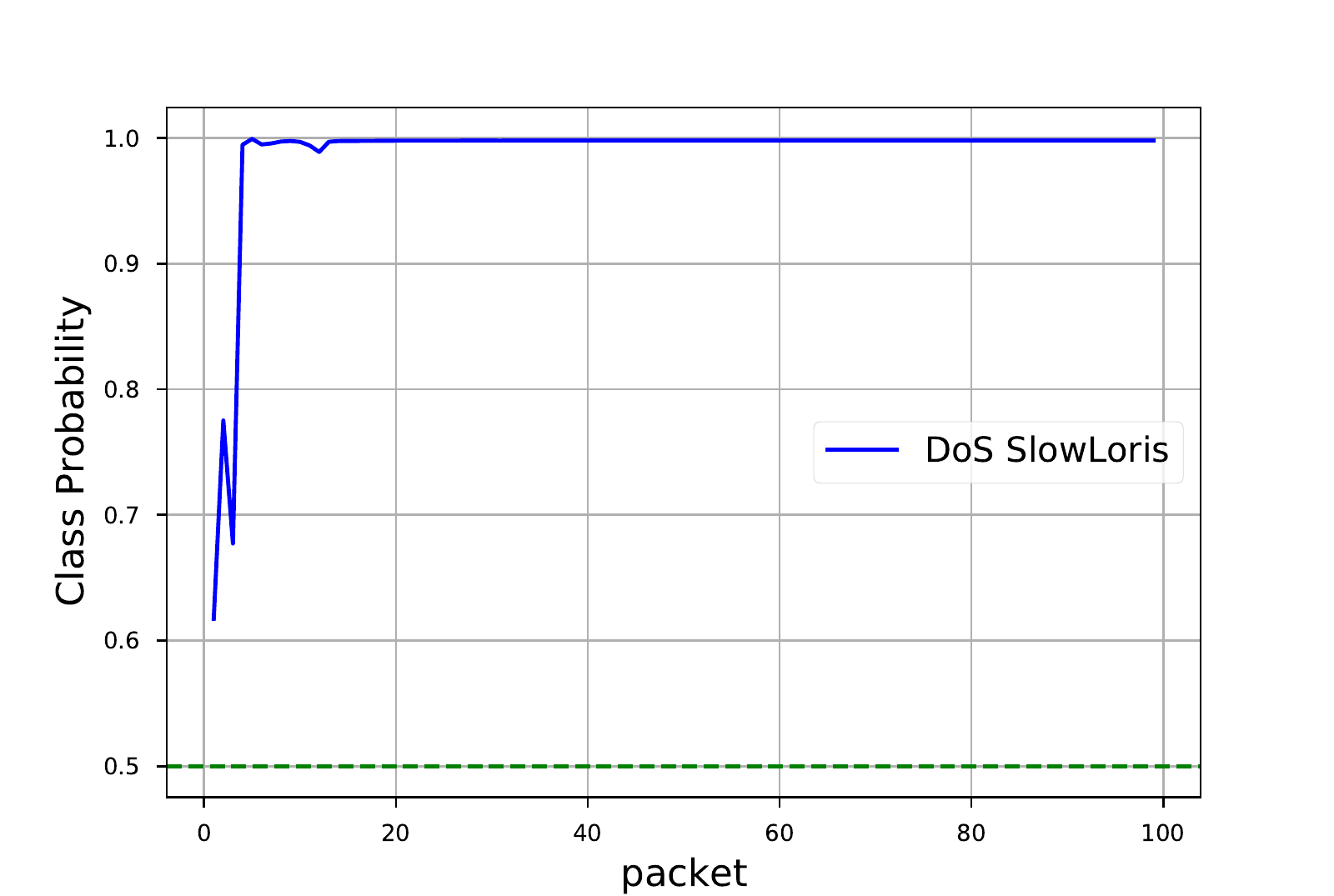}}
		\subfloat[DoS Hulk]{\includegraphics[width=.3\textwidth]{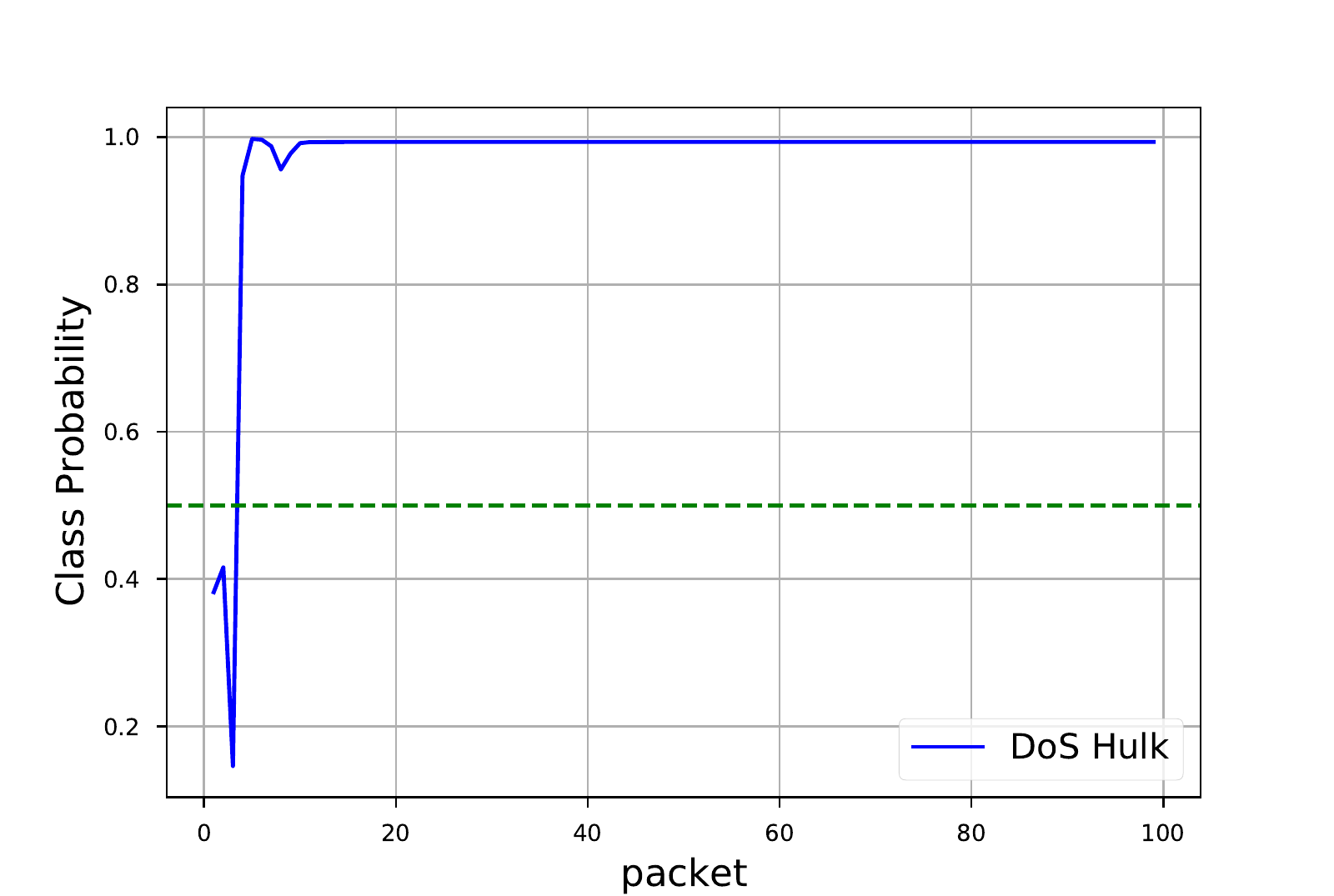}}
		\hfill
		\subfloat[DoS GoldenEye]{\includegraphics[width=.3\textwidth]{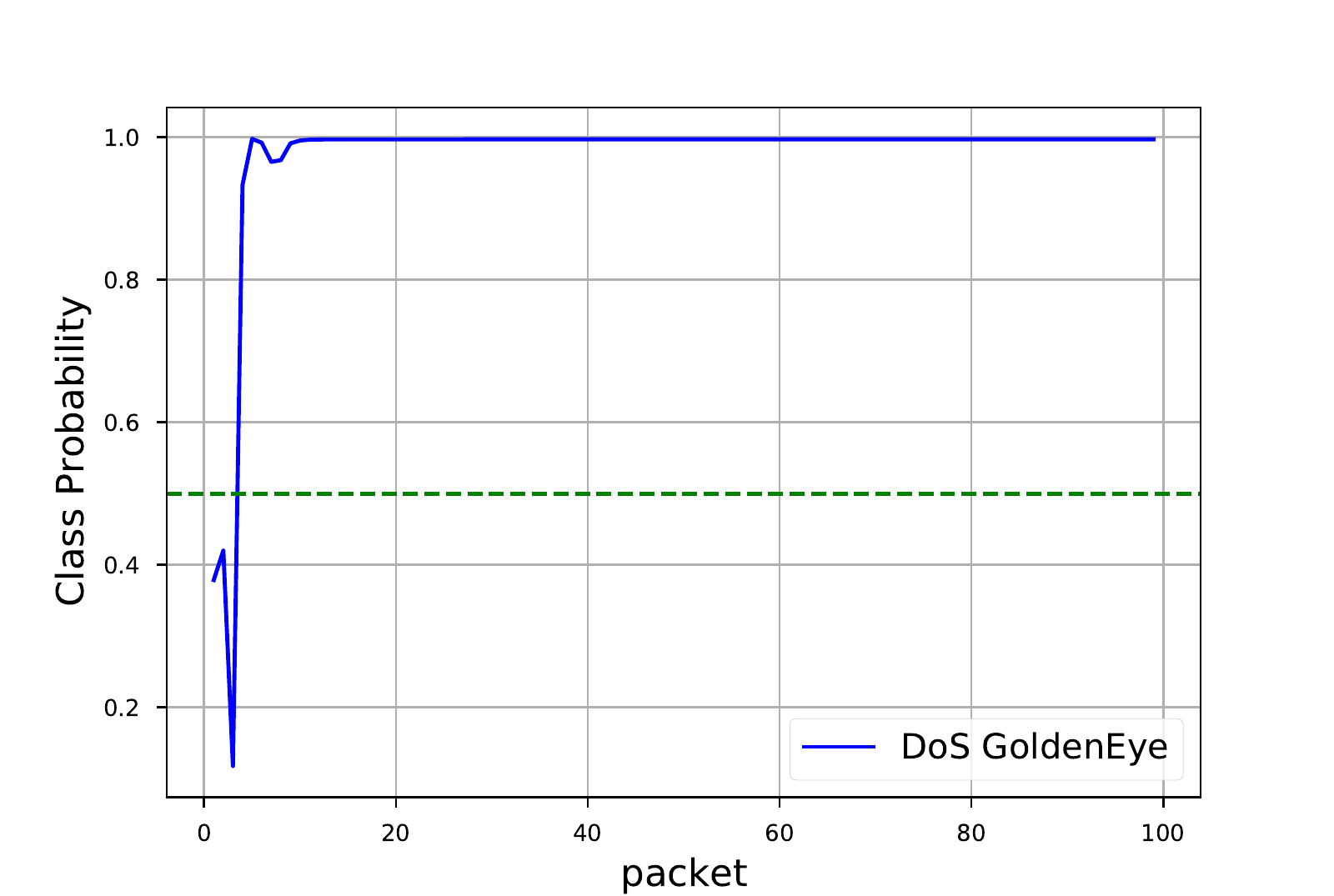}}
		\subfloat[FTP Patator]{\includegraphics[width=.3\textwidth]{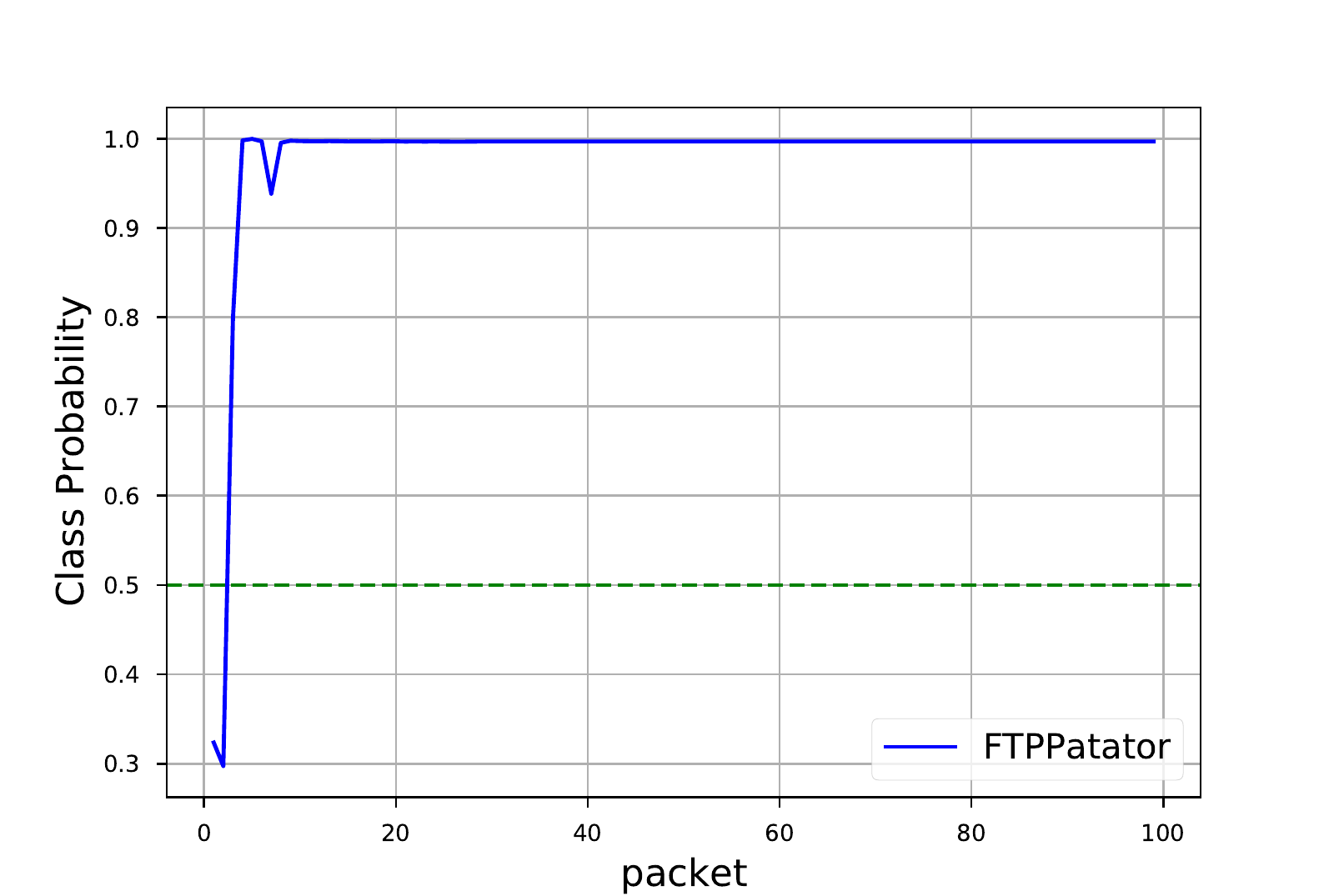}}
		\subfloat[SSH Patator]{\includegraphics[width=.3\textwidth]{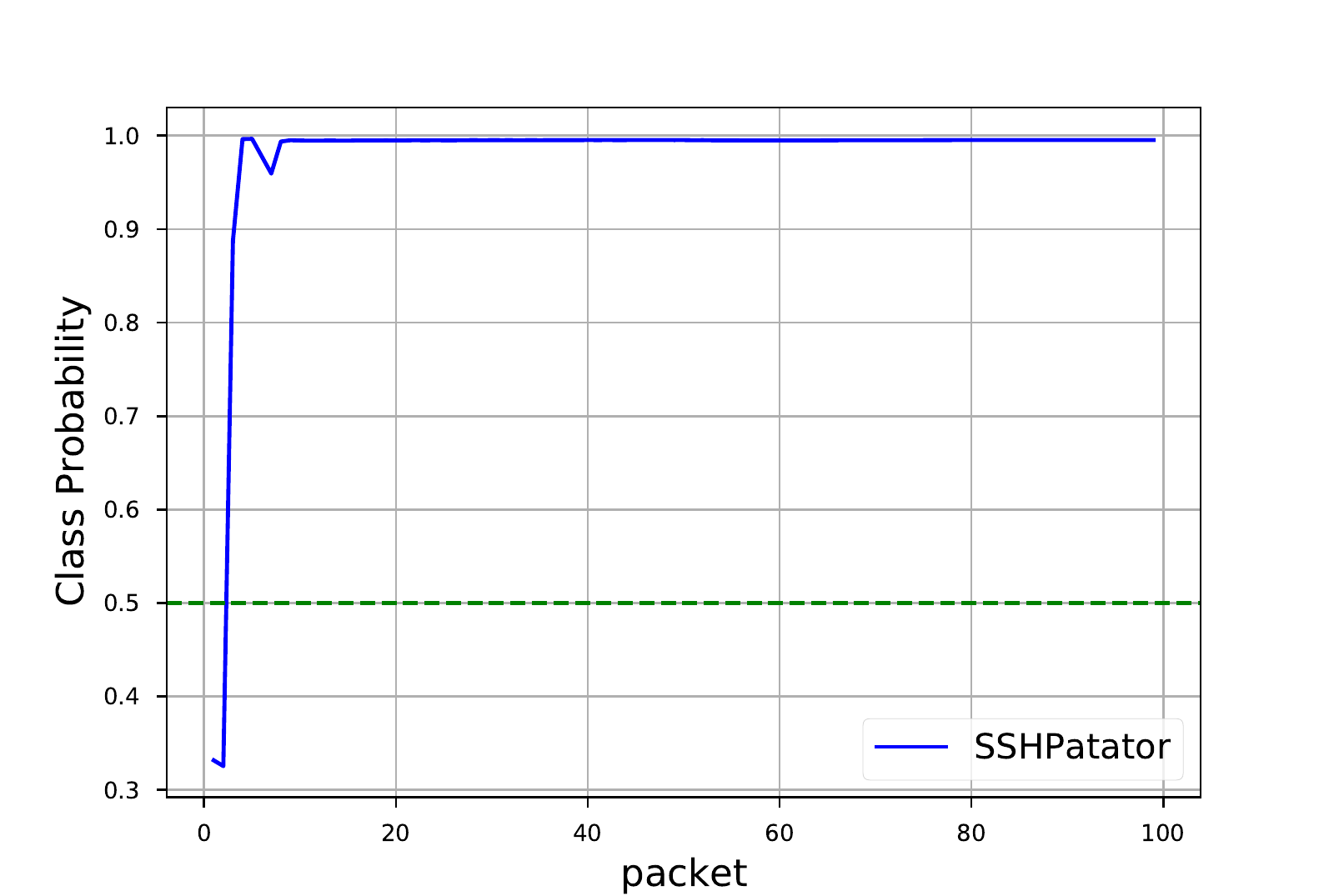}}
		\hfill
		\subfloat[BruteForce Web]{\includegraphics[width=.3\textwidth]{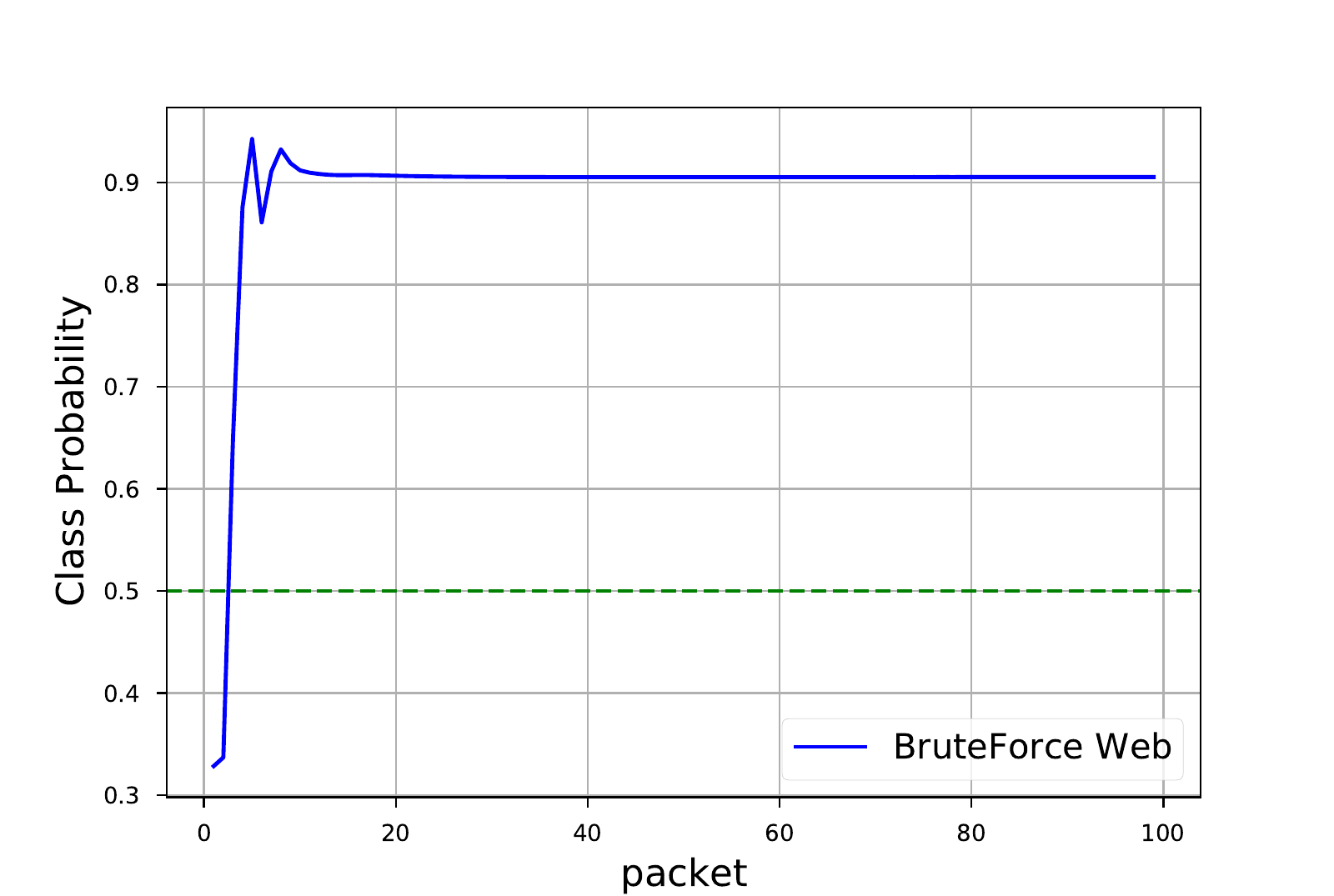}}
		\subfloat[BruteForce]{\includegraphics[width=.3\textwidth]{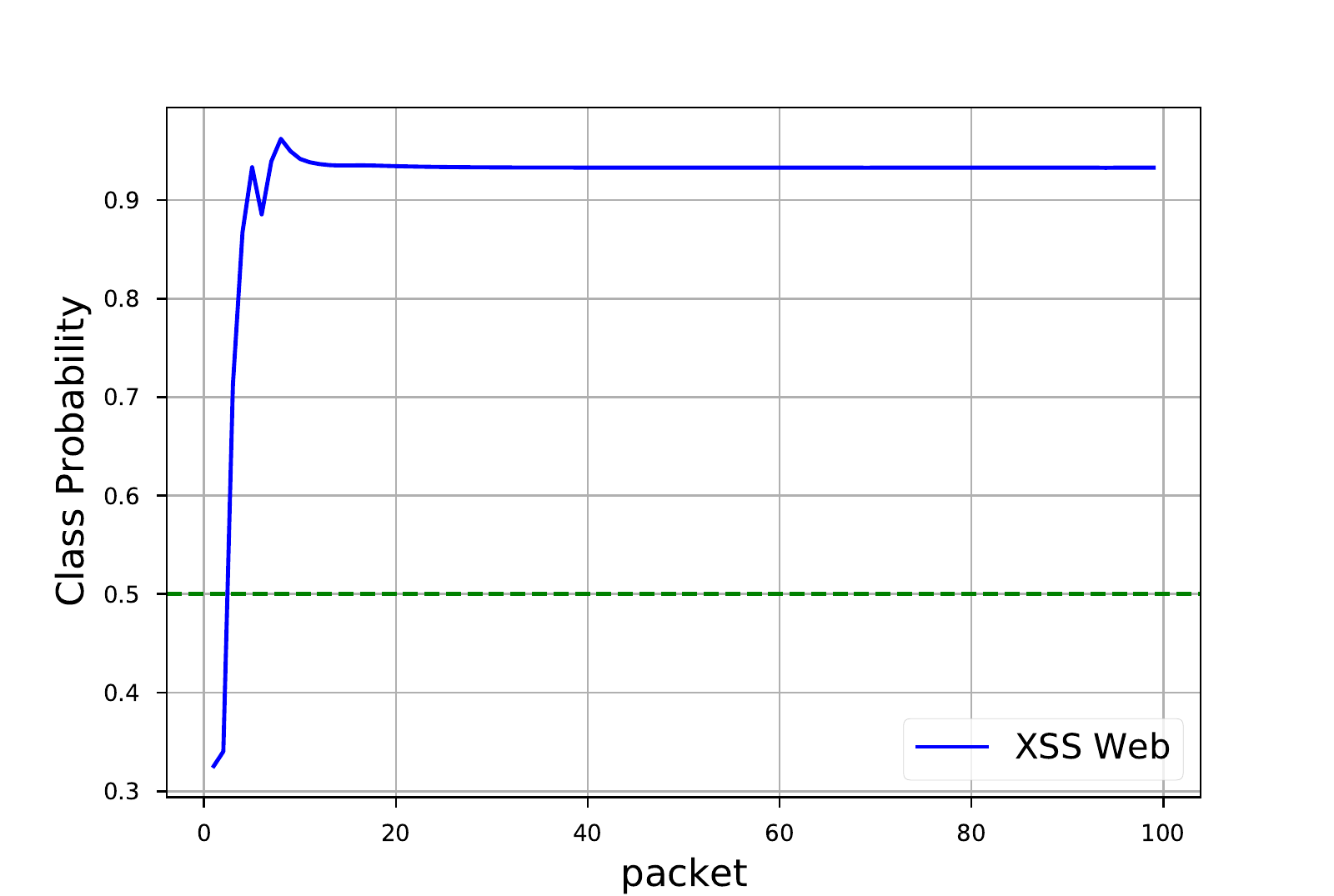}}
		\subfloat[Benign]{\includegraphics[width=.3\textwidth]{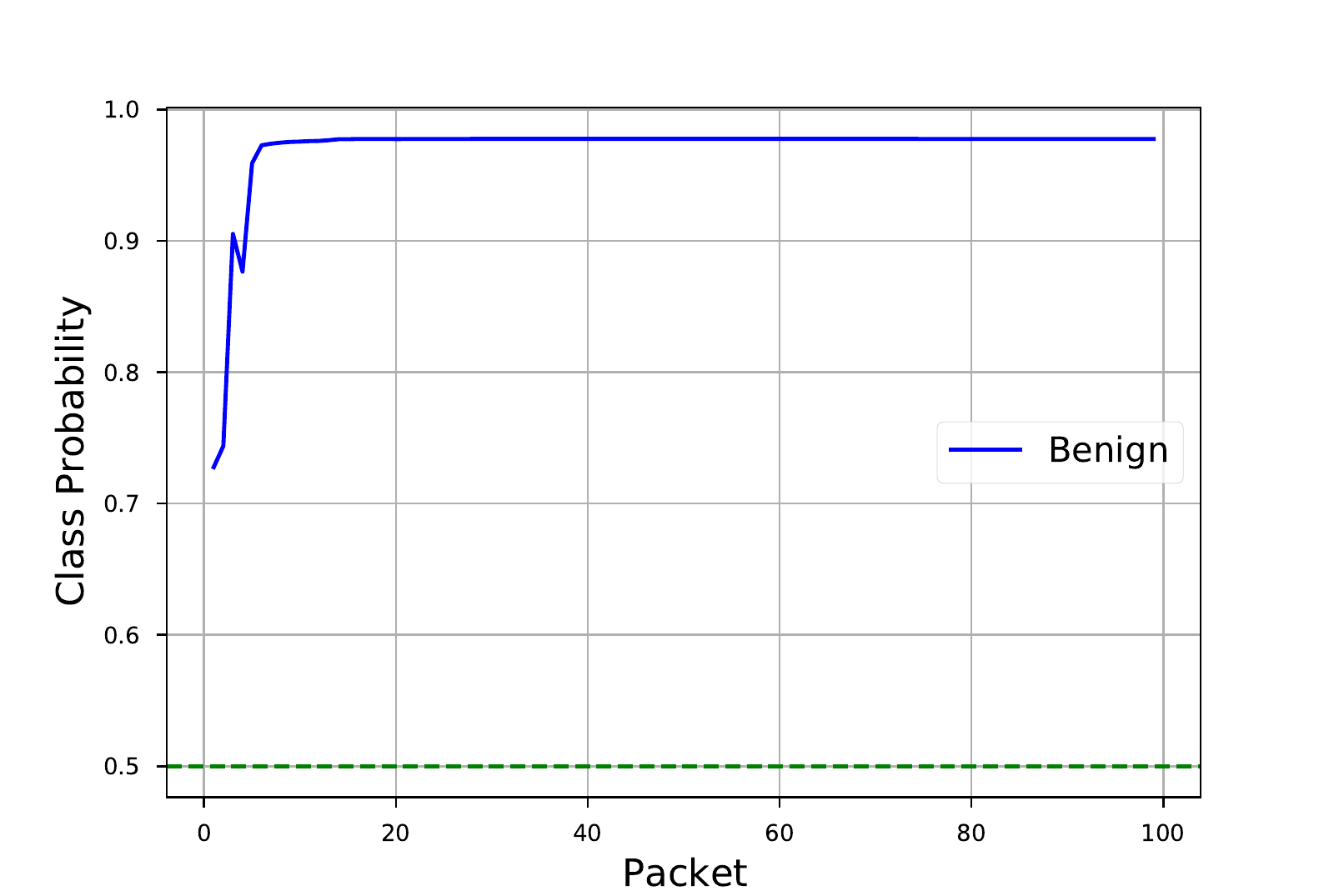}}
		\caption{The average true label probability of each flow's packet sequence in the CIC-IDS2017 dataset.}
		\label{fig:packet-labeling}
	\end{figure*}

	The average probability assigned by the model to the true (actual) label of a flow, as a function of each incoming packet is depicted in Figure \ref{fig:packet-labeling}. 
	The results demonstrate that with only 15 packets, the model can predict a flow's label with more than 80\% detection rate.

	\section{Discussion and Future Directions}\label{sec:discussion}

	
	The evaluations in Sections \ref{sec:DeepAdaptive} and \ref{sec:expfed} indicate that in terms of adaptability, CNN models tend to learn the new traffic patterns better than LSTM models.
	This phenomenon could be explained by the fact that CNN layers extract features at the flow level, which capture the spatial characteristics of packets in a given flow. 
	On the other hand, while LSTM layers are well-suited for obtaining the temporal relation between sequential packets, the feature vector extracted by them is based on the transferred history of the previous packets. Consequently, the direct data observation by CNN models can possibly generate better features for representing the flows. While the classification patterns based on these features might change over time (according to the traffic concept drift), those features themselves embody a suitable representation of a flow. Thus, the dense layers in CNN-based models have a more straightforward task for tuning their weights when facing new traffic. The weakness of LSTM models in the case of learning new attacks (Table \ref{LSTM:cont-2017}) is especially aggravated for attacks that use contents similar to benign flows (\eg, portscan and FTP Patator\footnote{Unlike FTP Patator, SSH Patator uses encrypted traffic. The randomness of the flow bytes makes it different from the benign traffic. Similarly, other attacks, such as web attacks and a variety of DOS attacks, use slightly different contents.}). 
	
	Furthermore, in our experiments, we have investigated the models' performances for adaptation to new traffic under strict constraints. To be more precise, the models are provided with a low amount of knowledge both at the initial training (\ie, only one known anomaly is used in the initial training phase) and updating phase (\ie, only 128 flows are used as the new traffic samples). According to our evaluations, by relaxing the above constraints, the results of LSTM-based models improve by training with more data.
	On the other hand, based on the results of Section \ref{sec:exppacket}, LSTM models can detect an anomaly with fewer packets, thus being more efficient and applicable to real-world scenarios. More precisely, the early detection capability of LSTM-based models can help mitigate the intrusion's impact on the target organization. Overall, the initial training of the LSTM-based models needs more effort, but they are more efficient both in detecting with less number of packets and the updating process (see Table~\ref{tab:CNN-vs-LSTM}).

	Regarding the catastrophic forgetting issue, the results in Section \ref{sec:DeepAdaptive} indicate that regardless of how well the model adapts itself to new traffic, its performance on its previous knowledge will not deteriorate. Figures \ref{fig:cnncont2017}, \ref{fig:cnncont2018}, \ref{fig:lstmcont2017}, and \ref{fig:lstmcont2018} indicate that after learning the new anomaly in each step, the model detection rate on previously learned anomalies is consistent with the previous step's detection rate on both new and old anomalies. 
	
	\begin{table}[!h]
		\caption{Resource and time consumption of CNN-based and LSTM-based architectures, where each number is averaged over different attacks.}
		\label{tab:CNN-vs-LSTM}
		\centering
		\resizebox{0.8\linewidth}{!}{
			\begin{tabular}{|c|l|l|l|l|l|l|}
				\hline
				\multicolumn{1}{|l|}{} & \multicolumn{1}{c|}{\parbox[c][1.2cm][c]{1.8cm}{\centering \textbf{Initial Model Size (Memory)}}} & \multicolumn{1}{c|}{\parbox[c][1.2cm][c]{1.8cm}{\centering \textbf{Expanded Model Size (Memory)}}} & \multicolumn{1}{c|}{\parbox[c][2cm][c]{2cm}{\centering \textbf{Initial Training (Time)}}} & \multicolumn{1}{c|}{\parbox[c][2cm][c]{2cm}{\centering \textbf{Updating/ Expansion (Time)}}} &
				\multicolumn{1}{c|}{\parbox[c][2cm][c]{2cm}{\centering \textbf{Updating/ Compression (Time)}}} & \multicolumn{1}{c|}{\parbox[c][2cm][c]{2cm}{\centering \textbf{Validation (Time)}}} \\ \hline
				\parbox[c][1.2cm][c]{2.3cm}{\centering \textbf{CNN-Based}}     & \parbox[c][0.4cm][c]{1.8cm}{\centering 300MB}                                              & \parbox[c][0.4cm][c]{1.8cm}{\centering 320MB}                                               & \parbox[c][0.4cm][c]{1.8cm}{\centering 7min}                                                    & \parbox[c][0.4cm][c]{1.8cm}{\centering 15s}  & \parbox[c][0.4cm][c]{1.8cm}{\centering 6min}                                           & \parbox[c][0.4cm][c]{1.8cm}{\centering 1.22s}                                         \\ \hline
				\parbox[c][1.2cm][c]{2.3cm}{\centering \textbf{LSTM-Based}}    & \parbox[c][0.4cm][c]{1.8cm}{\centering 20MB}                                               & \parbox[c][0.4cm][c]{1.8cm}{\centering 23MB}                                                & \parbox[c][0.4cm][c]{1.8cm}{\centering 13min}                                                   & \parbox[c][0.4cm][c]{1.8cm}{\centering 2s} & \parbox[c][0.4cm][c]{1.8cm}{\centering 2min}                                               & \parbox[c][0.4cm][c]{1.8cm}{\centering 2.97s}                                         \\ \hline
		\end{tabular}}
	\end{table}
	
	The IDS performance and its required resources are other determinative points in selecting the deep model architecture. According to Table \ref{tab:CNN-vs-LSTM}, LSTM-based models are more well-suited for practical IDSes. Although they need more time for the initial training of the model, they update themselves faster in continual updating procedures and consume less memory for their models. As mentioned, the reported initial training time (in Table \ref{tab:CNN-vs-LSTM}) is based on the average elapsed time for each of our different experiments with about 3000 to 5000 flows. The updating and validation times are reported according to processing 128 and 1000 flows, respectively.

	\begin{table*}[!ht]
		\caption{Comparison between the previous related studies and the proposed framework.}
		\label{discussion:comparison}
		\centering
		\resizebox{\linewidth}{!}{
			\begin{tabular}{|c|c|c|c|c|c|c|}
				\hline
				& \parbox[c][0.80cm][c]{1.5cm}{\centering \textbf{DL-based}} & \textbf{Unsupervised} & \parbox[c][0.4cm][c]{2cm}{\centering \textbf{Continuous \\ Adaptation}} & \textbf{Multi-Agent} & \parbox[c][0.4cm][c]{2cm}{\centering \textbf{Early Attack \\  Detection}} & \textbf{Dataset}                                                                             \\ \hline
				\textbf{C. Yin et al.   \cite{yin2017deep}}            & \checkmark        &                       &                                 &                      &                                   & \parbox[c][0.4cm][c]{5cm}{\centering NSL-KDD }                                                                                      \\ \hline
				\textbf{R. Vinayakumar et al. \cite{vinayakumar2017applying}}                   & \checkmark        &                       &                                 &                      &                                   &
				\parbox[c][0.4cm][c]{5cm}{\centering KDDCup 99 }                       \\ \hline
				\textbf{S. Thakur et al. \cite{thakur2021intrusion}}                   & \checkmark        &                       &                       &                      &                                   &

				\parbox[c][0.4cm][c]{5cm}{\centering CIC-IDS2017}                 \\ \hline
				\textbf{AM. Riyad et   al. \cite{riyad2019adaptive}}                 &        &             &   \checkmark                               &  \checkmark                      &                                   & 	\parbox[c][0.4cm][c]{5cm}{\centering KDD99}                                                                                    \\ \hline
				\textbf{C.   Kim et al. \cite{kim2019designing}}        & \checkmark        &             & \checkmark                      &                      &                                   & 
				\parbox[c][0.4cm][c]{5cm}{\centering KDD99, NSL-KDD}                                                                            \\ \hline
				\textbf{D.   Pamartzivanos et al. \cite{papamart2019}} & \checkmark                  & \checkmark            & \checkmark                      &                      &                                   & \parbox[c][0.4cm][c]{5cm}{\centering KDD99, NSL-KDD}                                                                                            \\ \hline
				\textbf{N. Gupta et al.   \cite{GUPTA2022102499}}            & \checkmark        &             &                                 &                      &                                   & \parbox[c][0.8cm][c]{5cm}{\centering NSL-KDD, CIDDS-001, CIC-IDS2017 }                                                                                      \\ \hline
				\textbf{Z. Wang et   al. \cite{WANG2021102177}}            & \checkmark        &             &                       &                      &                                   & \multicolumn{1}{c|}{\parbox[c][1cm][c]{7cm}{ \centering KDD99, NSL-KDD,\\ UNSW-NB15, CIDDS-001, ADFA-LD}}                                                                                            \\ \hline
				\textbf{Z. Wang et al. \cite{WANG2022102542}}    & \checkmark        &             &                                 &                      &                                   & \multicolumn{1}{c|}{\parbox[c][1cm][c]{7cm}{ \centering UNSW NS2019, ISCX IDS 2012, CIC-IDS2017, CIC-ANDMAL2017}}                                                                                         \\ \hline
				\textbf{G. F.   Cretu-Ciocarlie et al. \cite{Cretu2009anomaly}}    &         & \checkmark            & \checkmark                                &                      &                                   & \parbox[c][1cm][c]{8cm}{\centering Network Traffic of Columbia University's Computer Science Department }                                                                                         \\ \hline
				\textbf{F. Folino et al. \cite{FOLINO202148}}    & \checkmark        &  Semi-supervised           &      \checkmark                           &                      &                                   & \parbox[c][0.4cm][c]{5cm}{\centering CIC-IDS2017, ISCXIDS2012}                                                                                         \\ \hline
				\textbf{M. Soltani   et al. \cite{soltani2021adaptable}}    & \checkmark        &             &                                 &                      &                                   & \parbox[c][0.4cm][c]{5cm}{\centering CIC-IDS2017, CSE-CIC-IDS2018}                                                                                         \\ \hline
				\textbf{A. Mirza et al. \cite{mirza2018computer}}    & \checkmark        &             &                                 &                      &  \checkmark                                 & \parbox[c][0.4cm][c]{5cm}{\centering ISCXIDS2012}                                                                                        \\ \hline
				\textbf{J. Gao et al. \cite{gao2019lstm}}    & \checkmark        &             &                                 &                      &   \checkmark                                & \parbox[c][0.4cm][c]{5cm}{\centering  SCADA simulated testbed}                                                                                         \\ \hline
				\textbf{Proposed   Framework}                                   & \checkmark        &             & \checkmark                      & \checkmark           & \checkmark                        & \parbox[c][0.4cm][c]{5cm}{\centering CIC-IDS2017, CSE-CIC-IDS2018}               \\ \hline
		\end{tabular}}
	\end{table*}

	One should also consider the efficiency of the updating procedure in an adaptive deep intrusion detection system. An IDS should update itself with the traffic concept drift as early as possible. Consequently, in this paper, we evaluate the updating procedure (Section \ref{sec:DeepAdaptive}) with only 128 flows of the new traffic which is considered relatively low compared to the number of flows used to train an initial model (i.e., about 3000\textasciitilde5000 flows for each attack).
	
	Considering the distributed implementation of the proposed framework (evaluated in Section \ref{sec:expfed}), the federated distillation procedure yields acceptable results on both known and new anomalies while the agents learn novelty attacks and update the model asynchronously. As a result, the proposed multi-agent IDS framework can manage big data issue in practical situations. Furthermore, as discussed in Section~\ref{sec:framework_federated}, the proposed framework can also improve an agents' data privacy.

	In the following, to extend this research, we review possible directions for future studies.
	In the deep learning scope, it is observed that the adversarial attacks are a critical challenge for DL models \cite{9149117}\cite{MadryMSTV18}\cite{akhtar2018threat}. In these type of attacks, the model is misled with deceptive data. Consequently, in future studies, one can evaluate the proposed framework against adversarial attacks and devise defense solutions for reducing this threat.
	
	In the end, to complete our analysis, we compare the proposed framework with previous related research studies from different aspects. As demonstrated in Table \ref{discussion:comparison}, the proposed framework simultaneously provides solutions for the three aforementioned challenges of DL-based IDSes: continuous adaption, multi-agent IDSes, and early attack detection. Furthermore, note that most of the proposed DL-based IDS frameworks depend on labeled datasets. However, for practical applications, future studies can develop an unsupervised version of our proposed online adaptive anomaly detection framework. We believe that, in addition to the suggestions provided in this work, accomplishing this last step will result in a DL-based IDS more suitable for real-world scenarios.


	\section{Conclusion}\label{sec:conclusion}
	\label{conclusion}
	This paper presented a novel framework for DL-based IDSes that mitigates three practical issues these systems are currently facing. Namely, we provided solutions for continuously adapting the IDS to network concept drift, early attack detection, and efficiently functioning in a multi-agent environment (\eg, sharing the attack knowledge from different located IDS sensors, load-balancing the flows between different agents, and managing interleaving flows).
	
	The proposed framework exploits continual learning algorithms to update DL-based models for adapting to the concept drift in attack/benign traffic behaviors. Additionally, it uses federated learning for designing multi-agent IDSes and providing privacy and load balancing for big data traffics. Furthermore, the paper investigates the usage of Long Short-Term Memory networks (LSTMs) for packet labeling and early anomaly detection to design more practical IDSes. Finally, the framework is implemented and evaluated with two architectures: convolutional neural networks (CNNs) and LSTM-based models. The results indicate that while both architectures perform well, CNN models prevail in terms of detection rate, and LSTM models are more suitable for early anomaly detection with just a few packets. 
	
	
	%

	


	
	
	
	\bibliographystyle{elsarticle-num}
	\bibliography{references2}

\end{document}